\documentclass[12pt]{article}
\usepackage{jcappub}
\usepackage{bbm}
\usepackage{array}
\usepackage{float}
\usepackage{dsfont}
\usepackage{amstext}
\usepackage{psfrag}
\usepackage{a4}
\usepackage{a4wide}
\usepackage{hyperref}
\usepackage{multicol} 
\usepackage{multirow}
\usepackage{subcaption}
\usepackage{ulem}
\usepackage{xcolor,colortbl}
\usepackage{stackengine}
\usepackage{tabularx}
\usepackage{microtype}
\definecolor{darkteal}{RGB}{0,114,153}
\definecolor{teal}{RGB}{0,125,114}
\allowdisplaybreaks[1]
\newcommand\xrowht[2][0]{\addstackgap[.5\dimexpr#2\relax]{\vphantom{#1}}}

\newcommand{\leftparbox}[2]{\parbox{#1}{\begin{flushleft} #2 \end{flushleft}}}
\newcommand{\twotwosig}[4][\pbwidth]{
	\begin{equation}
		\left.
		\begin{aligned}
			#2 \\ #3
		\end{aligned}
		\ \right\} \ \ \mbox{\text{\leftparbox{#1}{95\,\%,~#4}}}
	\end{equation}}

\newcolumntype{g}{>{\columncolor{Gray}}c}
\newcolumntype{w}{>{\columncolor{Gray2}}c}
\definecolor{Gray}{gray}{0.90}
\definecolor{Gray2}{gray}{0.98}
\title{Smooth $\mu$-Hybrid and Non-Minimal Higgs Inflation in $SU(4)_{C}\times SU(2)_{L}\times SU(2)_{R}$ With Observable Gravitational Waves}

\author{Umer Zubair}

\affiliation{Department of Physics and Astronomy, \\ University of Delaware, Newark, DE 19716, USA}

\emailAdd{umer@udel.edu}

\abstract{We propose to study a smooth variant of the $\mu$-hybrid inflation model and a non-minimal Higgs model of inflation with quartic non-minimal coupling between the Higgs field and gravity within the context of a realistic GUT gauge group based on supersymmetric $SU(4)_{C}\times SU(2)_{L}\times SU(2)_{R}$. These models are incorporated with a realistic scenario of reheating and non-thermal leptogenesis, compatible with the constraints from the baryon asymmetry of the universe. Notably, both models successfully address the MSSM $\mu$-problem and avoid the issue of primordial magnetic monopoles. Our analysis reveals that both models predict a scalar spectral index $n_s$ that closely aligns with the central observationally favored value of Planck2018 + BICEP2/Keck Array (BK15) data and yield a large tensor-to-scalar ratio ($r > 10^{-3}$), potentially detectable in forthcoming CMB experiments.
}

\begin{document}
	\maketitle
	
	\section{Introduction}
	
	Observations thus far have confirmed several key predictions of the most basic inflationary models. These include a spectrum that is nearly, though not precisely, scale-invariant, consisting of predominantly Gaussian and adiabatic scalar perturbations \cite{Planck:2018jri}. Within the framework of inflationary theory, the mechanism driving the formation of cosmic structures may also generate a significant background of primordial gravitational waves (PGWs). These waves, referred to as tensor perturbations, imprint their signature on the polarization of the Cosmic Microwave Background (CMB) \cite{Kamionkowski:2015yta}, offering a potential avenue for observation using present-day technology \cite{Seljak:1996gy, Kamionkowski:1996ks, Seljak:1996ti}. A detection of the tensor signal would offer unprecedented insight into the earliest epochs of our Universe, bolstering the case for large-field inflation and offering profound implications for our comprehension of the inflationary period and physics at extremely high energies \cite{Baumann:2014nda}.
	
	Given that the energy scale of inflation closely aligns with SUSY gauge unification, SUSY hybrid inflation \cite{dvali:1994, Linde:1997, Senoguz:2003, Senoguz:2005, Copeland:1994ej, Buchmuller2000183, REHMAN:2010191, REHMAN:201075} naturally integrates into grand unified theories, with the GUT scale \( M_{\text{GUT}} \) providing the correct order of magnitude for the amplitude of primordial scalar fluctuations \cite{dvali:1994}. Thus, supersymmetric hybrid inflation serves as an excellent framework for bridging inflationary cosmology with particle physics at the GUT scale. On the other hand, there is also a compelling temptation to incorporate Higgs inflation into SUSY grand unified theories within a supergravity framework. 
	
	In the minimal SUSY hybrid inflation model \cite{Linde:1997, REHMAN:2010191, REHMAN:201075}, inflation ends abruptly, transitioning into a 'waterfall' phase where undesirable topological defects, such as magnetic monopoles, can arise \cite{R.Jeannerot_2001} if predicted by the GUT gauge group, potentially leading to cosmological catastrophes. To address this issue, various extensions to inflation models have been proposed, including shifted hybrid inflation \cite{Jeannerot_2000, khalil:2021, zubair:2023q1}, smooth hybrid inflation \cite{Lazarides:1995, mansoor:2012, mansoor:2015}, and new inflation \cite{IZAWA1997331, IZAWA1997249, mansoor:2020, mansoor2023108}. Among these, the smooth hybrid inflation model stands out for several reasons. Unlike standard and shifted hybrid inflation, the potential of smooth hybrid inflation possesses a slope at the classical level, driving inflation without the need for radiative corrections. Moreover, the system follows a specific valley of minima, leading to a particular point on the vacuum manifold, thereby preventing the formation of topological defects at the end of inflation, a feature shared also by shifted hybrid inflation.
	
	The non-minimal Higgs inflation \cite{mansoor:2021ma, C.Pallis_2011} emerges from a non-minimal coupling between the Inflaton Higgs field and the Ricci scalar, $\mathcal{R}$. The GUT gauge group spontaneously breaks during inflation due to the non-zero values acquired by the relevant Higgs superfields, thereby avoiding the problem of primordial magnetic monopoles. Additionally, akin to the smooth hybrid inflation model, the potential in non-minimal Higgs inflation exhibits inclination at the tree-level, obviating the necessity for radiative corrections.
	
	The Pati-Salam symmetry $G_{422}$ exhibits numerous remarkable features and is an attractive choice for a grand unified theory (GUT). It provides a natural framework for implementing the leptogenesis scenario due to the presence of right-handed neutrinos, leading to tiny but non-zero neutrino masses through the seesaw mechanism. Furthermore, it facilitates third-family Yukawa unification (YU) \cite{rizwan:2019mh, Gogoladze:2009ks, ahmed:2023uz, Gogoladze:2011ksr, Gogoladze:2010ksr, Raza:2015su}, charge quantization, and avoids the doublet-triplet splitting problem since both Higgs doublets are contained in a bi-doublet rather than the GUT scale Higgs fields. The combination of third-family Yukawa unification with non-universalities in the gaugino or the scalar sector can be made consistent with low-energy phenomenology. For the amelioration of the little hierarchy problem in $G_{422}$ model with the help of non-universal gaugino masses at the GUT scale, see \cite{mansoor:2009sg}. Additionally, due to the $U(1)_R$ symmetry, various dangerous proton decay operators are effectively suppressed. The unbroken $\mathbb{Z}_2^{\text{mp}} \subset U(1)_R$ acts as matter parity, implying the existence of a stable lightest supersymmetric particle (LSP) and thus a potential dark matter candidate.
	
	In this paper, we introduce the smooth $\mu$-hybrid inflation model for the first time, employing the mechanism proposed in \cite{dvali_1998259} for generating the MSSM $\mu$-term in combination with smooth hybrid inflation. This model is embedded in a realistic GUT based on $SU(4)_{C}\times SU(2)_{L}\times SU(2)_{R}$ gauge symmetry and possesses several notable features. It naturally avoids the magnetic monopole problem, generates the MSSM $\mu$-term, and does not require radiative corrections due to the inclination present at the tree level. This is in contrast to the shifted $\mu$-hybrid inflation \cite{mansoor:vardag2021, mansoor:adeela2023}, where one-loop radiative corrections play an important role in driving inflation. Additionally, using the same superpotential, we study a non-minimal Higgs inflation model by employing a special form of Kahler potential \cite{mansoor:201928}, resulting in a quartic non-minimal coupling of the inflaton Higgs field to gravity. In this setup, the conjugate Higgs fields assume the role of the inflaton, driving inflation. Both models incorporate a realistic scenario of reheating and non-thermal leptogenesis consistent with the constraints from the observed baryon asymmetry of the universe (BAO). The predictions of both models for the scalar spectral index $n_s$ are in accordance with the Planck2018 + BICEP2/Keck Array (BK15) data \cite{Planck:2018jri}, and they readily accommodate observable values of the tensor-to-scalar ratio $r$, which can be probed by a variety of forthcoming CMB experiments. Lastly, the proton decay rate is heavily suppressed to have any observable signature in the next generation of experiments, rendering it practically stable.
		
	The remainder of the paper is structured as follows: Section \ref{sec2} presents an overview of the supersymmetric $SU(4)_{C}\times SU(2)_{L}\times SU(2)_{R}$ model. Section \ref{sec3} delves into the implementation of smooth $\mu$ hybrid inflation, along with a discussion on the inflationary predictions and their comparison with observational data. Section \ref{sec4} focuses on the implementation of the non-minimal Higgs inflation model, presenting results, predictions and their comparison with observations. The potential detection of observable primordial gravitational waves in upcoming experiments is briefly addressed in Section \ref{sec5}, while Section \ref{sec6} provides a brief discussion on proton decay. Finally, the findings are summarized in Section \ref{sec7}.
	
	\section{The SUSY $SU(4)_{C}\times SU(2)_{L}\times SU(2)_{R}$ Model} \label{sec2}%
	
	The SUSY Pati-Salam (PS) model is based on the gauge symmetry $G_{422} = SU(4)_{C}\times SU(2)_{L}\times SU(2)_{R}$. The MSSM matter superfields, including the right handed neutrino, reside in the following irreducible representations of $G_{422}$ \cite{King:1997ia},
	\begin{align}
		F_i &=(\mathbf{4}, \mathbf{2},\mathbf{1})\equiv
		\left( {\begin{array}{cccc}
				u_{ir} &  u_{ig}  &  u_{ib}  & \nu_{il} \\
				d_{ir} &  d_{ig}  &  d_{ib}  & e_{il} \\
		\end{array} } \right),\nonumber \\
		F^c_i &=(\bar{\mathbf{4}}, \mathbf{1}, \mathbf{2})\equiv 
		\left( {\begin{array}{cccc}
				u^c_{ir} &  u^c_{ig}  &  u^c_{ib}  & \nu^c_{il} \\
				d^c_{ir} &  d^c_{ig}  &  d^c_{ib}  & e^c_{il} \\
		\end{array} } \right),
	\end{align}
	where $i=1,2,3$ denotes the family index and the subscripts $r$, $g$, $b$, $l$ denote the four colors of $SU(4)_C$. The model unifies each family of quarks and leptons into above two representations. Note that $F$ and $F^c$ comprise the $\mathbf{16}$ (spinorial) representation of $SO(10)$: $\mathbf{16} \rightarrow F (\mathbf{4}, \mathbf{2},\mathbf{1}) + F^c (\bar{\mathbf{4}}, \mathbf{1}, \mathbf{2})$. 
	
	The Higgs sector of $G_{422}$ consists of the following superfields: \textbf{(i)} a pair of GUT Higgs superfields $H^c$ and $\bar{H}^c$ represented as,
	\begin{align}
		H^c&=(\bar{\mathbf{4}}, \mathbf{1}, \mathbf{2})\equiv 
		\left( {\begin{array}{cccc}
				u^c_{Hr} &  u^c_{Hg}  &  u^c_{Hb}  & \nu^c_{Hl}\\
				d^c_{Hr} &  d^c_{Hg}  &  d^c_{Hb}  & e^c_{Hl} \\
		\end{array} } \right),\nonumber \\
		\bar{H}^c &=(\mathbf{4}, \mathbf{1}, \mathbf{2})\equiv 
		\left( {\begin{array}{cccc}
				\bar{u}^c_{Hr} &  \bar{u}^c_{Hg}  & \bar{u}^c_{Hb}  & \bar{\nu}^c_{Hl}\\\
				\bar{d}^c_{Hr} & \bar{d}^c_{Hg}  & \bar{d}^c_{Hb}  &\bar{e}^c_{Hl} \\
		\end{array} } \right),
	\end{align}
	which trigger the spontaneous breaking of $G_{422}$ gauge symmetry to $G_{\text{SM}}$ by acquiring nonzero vacuum expectation values (VEVs) along the right-handed sneutrino directions ($\lvert \langle \nu_{Hl}^c \rangle \rvert = \lvert \langle \bar{\nu}^c_{Hl} \rangle \rvert = M $) with  the following symmetry breaking pattern
	\begin{equation*}
		SU(4)_{C}\times SU(2)_{L}\times SU(2)_{R} \xrightarrow[M]{\langle H^c, \bar{H}^c \rangle} SU(3)_C \times SU(2)_L \times U(1)_{Y}\, ;
	\end{equation*}
	\textbf{(ii)} a bi-doublet Higgs superfield $h$ represented as,
	\begin{equation}
		h=(\mathbf{1}, \mathbf{2}, \mathbf{2})\equiv (h_u \ \ h_d)= \left( {\begin{array}{cc}
				h^+_u &   h^0_d   \\
				h^0_u &     h^-_d   \\
		\end{array} } \right),\\
	\end{equation}
	which, after the $G_{422}$ symmetry breaking, splits into electroweak Higgs doublets $h_u$ and $h_d$ whose neutral components subsequently develop VEVs $\langle h_u \rangle = \upsilon_u$, $\langle h_d \rangle = \upsilon_d$ with $\tan \beta = \upsilon_d / \upsilon_u$; \textbf{(iii)} a sextet Higgs superfield $G = (\mathbf{6}, \mathbf{1}, \mathbf{1})$, which after $G_{422}$ symmetry breaking, splits into two color-triplet Higgs superfields $g$, $g^c$ providing superheavy masses to the color-triplet pair $d^c_H$ and $\bar{d}^c_H$; \textbf{(iv)} and finally, a gauge singlet superfield $S = (\mathbf{1}, \mathbf{1}, \mathbf{1})$ whose scalar component acts as an inflaton. The bi-doublet $h$ and the sextet $G$ comprise the $\mathbf{10}$ representation of $SO(10)$: $\mathbf{10} \rightarrow h (\mathbf{1}, \mathbf{2}, \mathbf{2}) + G (\mathbf{6}, \mathbf{1}, \mathbf{1})$. The decomposition of the above $G_{422}$ representations under the SM gauge group, along with their global $U(1)_R$ and $Z_2$ charges, is given in Table \ref{tab:field_charges}.
	
	\begin{table}[t]
		\centering
		\begin{tabular}{cccc}
			\hline \hline  \rowcolor{Gray}\xrowht{10pt}
			&       \multicolumn{2}{c}{\textsc{Global Symmetries}}           &                    \\
			\rowcolor{Gray}
			\multirow{-2}{*}{\begin{tabular}[c]{@{}c@{}}$G_{422}$ \\\textsc{Representations}\end{tabular}} & \quad $U(1)_R$            & $\mathbb{Z}_2$  & \multirow{-2}{*}{\begin{tabular}[c]{@{}c@{}}\textsc{Decomposition}\\ \textsc{under} $G_{\text{SM}}$\end{tabular}}                    \\
			\hline \hline \rowcolor{Gray2} \multicolumn{4}{c}{\xrowht{10pt}\textsc{Matter sector}}\\
			\hline
			\xrowht{10pt}
			$F_i\ (\mathbf{4},\ \mathbf{2},\ \mathbf{1})$ &1 & $1$  & $Q_{i}\ (\mathbf{3},  \mathbf{2},   1/6)$\\
			\xrowht{10pt}
			&   &    &$L_i\ (\mathbf{1},  \mathbf{2},  -1/2)$  \\[10pt]
			\xrowht{10pt}
			$F^c_i\ (\bar{\mathbf{4}},\ \mathbf{1},\ \mathbf{2})$ & 1 & 1 & $u^c_{i}\ (\bar{\mathbf{3}},  \mathbf{1},  -2/3)$\\
			\xrowht{10pt}
			&  &  &  $d^c_{i}\ ( \bar{\mathbf{3}},  \mathbf{1},  1/3)$\\
			\xrowht{10pt}
			& &  &$\nu^c_i \ ( \mathbf{1},  \mathbf{1},  0)$ \\
			\xrowht{10pt}
			&  &  &  $e^c_i \ (\mathbf{1},  \mathbf{1},  1)$\\
			\hline \rowcolor{Gray2} \multicolumn{4}{c}{\xrowht{10pt}\textsc{Higgs sector}}\\
			\hline
			\xrowht{10pt}
			$H^c \ ({\bar{\mathbf{4}},\ \mathbf{1},\ \mathbf{2}})$ & 0 & 1 & $u^c_{H}\ ({ \bar{\mathbf{3}},  \mathbf{1}},  -2/3)$\\
			\xrowht{10pt}
			&  &  &  $d^c_{H}\ ({ \bar{\mathbf{3}},  \mathbf{1}},  1/3)$\\
			\xrowht{10pt}
			&  &  & $\nu^c_H \ ({ \mathbf{1},  \mathbf{1}},   0)$\\
			\xrowht{10pt}
			&  &   &$e^c_H \ ({ \mathbf{1},  \mathbf{1}},  1)$ \\[10pt]
			\xrowht{10pt}
			$\bar{H}^c \ ({\mathbf{4},\ \mathbf{1},\ \mathbf{2}})$ &0  & $-1$ & $\bar{u}^c_{H}\ ({ \mathbf{3},  \mathbf{1}},  2/3)$\\
			\xrowht{10pt}
			&  &  & $\bar{d}^c_{H}\ ({\mathbf{3},  \mathbf{1}},  -1/3)$\\
			\xrowht{10pt}
			&  &  & $\bar{\nu}^c_H \ ({ \mathbf{1},  \mathbf{1}},   0)$ \\
			\xrowht{10pt}
			&  &  &$\bar{e}^c_H \ ({ \mathbf{1},  \mathbf{1}},   -1)$\\[10pt]
			\xrowht{10pt}
			$h\ ({\mathbf{1},\ \mathbf{2},\ \mathbf{2}})$ &0 & $1$  & $h_u \ ({\mathbf{1},  \mathbf{2}},  1/2)$\\
			\xrowht{10pt}
			&  &  &$h_d\ ({ \mathbf{1},\ \mathbf{2}},\ -1/2)$ \\[10pt]
			\xrowht{10pt}
			$ G \ ({\mathbf{6},\ \mathbf{1},\ \mathbf{1}})$ &2 & $1$  & $g\ ({\mathbf{3},  \mathbf{1}},  -1/3)  $\\
			\xrowht{10pt}
			&  &  &$ g^c\ ({\bar{\mathbf{3}},  \mathbf{1}},  1/3)$ \\[10pt]
			\xrowht{10pt}
			$ S \ ({\mathbf{1},\ \mathbf{1},\ \mathbf{1}})$ &2 & $1$ & $ S\ ({ \mathbf{1},  \mathbf{1}},  0)$\\
			\hline \hline
		\end{tabular}
		\caption{The representations of matter and Higgs superfields under $G_{422}$ gauge symmetry, their decomposition under $G_{\text{SM}}$ along with their global $U(1)_R$ and $Z_2$ charges.}
		\label{tab:field_charges}
	\end{table}
	
	The superpotential of the model consistent with the $U(1)_R$ and $G_{\text{422}}$ symmetries is given by \cite{mansoor:2020pd422}
	\begin{align}\label{SP1}
		W & = \kappa S \left(M_{*}^2 - \frac {(\bar{H}^c H^c)^2}{m_P^2} \right)+\lambda Sh^2 \nonumber \\ 
		&+ \gamma^{ij}F^c_iF_jh + a\,G H^c H^c + b\,G \bar{H}^c \bar{H}^c \nonumber  \\ 
		&+ \frac{\bar{H}^c \bar{H}^c}{m_P} \left(\alpha^{ij}_1 F^c_iF^c_j + \alpha^{ij}_2 F_iF_j\right) +  \frac{{H}^c {H}^c}{m_P} \left(\beta^{ij}_1 F^c_iF^c_j + \beta^{ij}_2 F_iF_j\right) ,
	\end{align}
	where $\kappa$, $\lambda$, $\gamma^{ij}$, $\alpha^{ij}_{1,2}$, $\beta^{ij}_{1,2}$, $a$ and $b$ are real and positive dimensionless couplings and $M_{*}$ is a super heavy mass scale. The UV cutoff $m_P$ (reduced Planck mass) has replaced the cutoff $\Lambda$ \cite{mansoor:2012} typically employed in smooth hybrid inflation models \cite{Lazarides:1995}, which controls the non-renormalizable terms in the superpotential. The global $U(1)_{R}$ symmetry plays essential role in realizing successful inflation. It ensures the linearity of superpotential $W$ in $S$, forbidding non-linear terms which could spoil inflation \cite{dvali:1994}. Its unbroken $\mathbb{Z}_2^{\text{mp}}$ subgroup acts as ‘matter parity’, which implies a stable LSP, thereby making it a plausible dark matter candidate. Finally, it forbids several dangerous proton decay operators. The superpotential in \eqref{SP1} respects an additional discrete $\mathbb{Z}_2$ symmetry under which the combination $H^c \bar{H}^c$ is odd, and therefore, only even powers of the combination $H^c \bar{H}^c$ are allowed. All other fields are neutral under this $\mathbb{Z}_2$ symmetry. Note that the terms $(H^c)^4$ and $(\bar{H}^c)^4$ can also appear in the superpotential but become irrelevant in the $D$-flat direction. 
	
	The first two terms in the first line of the superpotential \eqref{SP1} are pertinent to smooth hybrid inflation, while the third term ($\lambda S h_u h_d$) gives rise to the effective $\mu$-term, derived below. The first term in the second line encompasses the Yukawa couplings, providing masses to quarks and leptons following electroweak symmetry breaking. The last two terms, involving the sextuplet superfield $G$, yield superheavy masses for the color-triplets $d^c_H$ and $\bar{d}^c_H$. The terms in the third line are relevant for proton decay while the term with coupling $\alpha_1^{ij}$ yields heavy right-handed neutrino masses for the see-saw mechanism.
	
	\section{Smooth $\mu$-Hybrid Inflation}\label{sec3}
	
	The superpotential terms relevant for smooth $\mu$-hybrid inflation are
	\begin{equation}
		W \supset \kappa S \left(M_{*}^2 - \frac {(\bar{H}^c H^c)^2}{m_P^2} \right)+\lambda Sh^2,
	\end{equation}
	where $h^2$ denotes the unique bilinear invariant $\epsilon_{ij} h_u^i h_d^j$. The global SUSY scalar potential obtained from the above superpotential is given by
	\begin{eqnarray}
		V &=& \kappa^2\, \Big| M_{*}^2 - \frac {(\bar{H}^c H^c)^2}{m_P^2} + \frac{\lambda}{\kappa} h^2 \Big|^2 + 4 \lambda^2 h^2 \left| S \right|^2 \nonumber \\
		&+& \frac{4 \, \kappa^2 \lvert S \rvert^2 \, \lvert\bar{H}^c\rvert^2 \, \lvert H^c\rvert^2}{m_P^4} \left(\lvert\bar{H}^c\rvert^2 + \lvert H^c\rvert^2\right) + D\text{-terms},
		\label{scalarpot-smooth}
	\end{eqnarray}
	where the scalar components of the superfields are denoted by the same symbols as the corresponding superfields. The VEV’s of the fields at the global SUSY minimum of the above potential are given by,
	\begin{gather}
		\langle S \rangle = 0, \quad  \langle h \rangle = 0, \quad \quad  \langle \bar{H}^c H^c \rangle = M^2 = M_{*} m_P.
		\label{gmin}
	\end{gather} 
	\begin{figure}[t]
		\centering 
		\includegraphics[width=8cm]{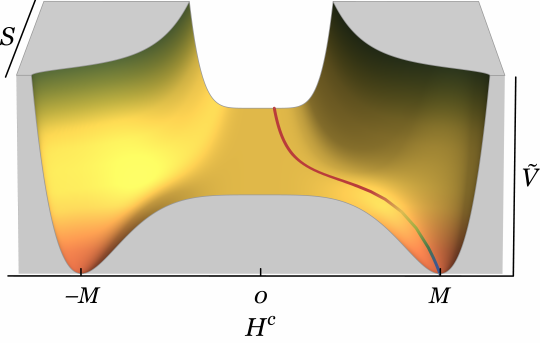}
		\caption{The tree-level, global SUSY scalar potential ($\tilde{V} = V/V_0$) features two symmetric valleys of local minima that can serve as inflationary trajectories. These valleys, while not classically flat, already exhibit an inclination at the tree level, can drive the inflaton towards the SUSY vacua. Unlike standard SUSY or shifted hybrid inflation scenarios, there is no need for radiative corrections, which typically contribute subdominantly to the slope of the inflationary trajectories.}
		\label{fig:pot_plot}
	\end{figure}
	The $D$-flatness condition yields $\bar{H}^{c*} = e^{\dot{\iota}\theta} H^c$ and $h_{ui} = e^{\dot{\iota}\phi} \epsilon_{ij} h_d^{j*}$, where $\theta$ and $\phi$ are arbitrary phases. Restricting ourselves to the direction with $\theta = 0$ ($\bar{H}^{c} =  H^c$), which contains the smooth inflationary path and the SUSY vacua, as well as stabilizing the potential at $h = 0$, we obtain,
	\begin{equation}
		V =  \kappa^2 M_{*}^4 \left[\left(1 - \frac {\lvert H^c\rvert^4}{M^4} \right)^2 + 8 \frac{\lvert S \rvert^2 \, \lvert H^c\rvert^6}{M^8}\right] ,
		\label{scalarpot-smooth2}
	\end{equation}
	The above scalar potential can be written in terms of the dimensionless variables
	\begin{equation}
		y = \frac{\lvert H^c\rvert}{M}, \qquad z = \frac{\lvert S \rvert}{M},
	\end{equation}
	as follows,
	\begin{equation}
		V =  V_0 \left(\left(1 - y^4 \right)^2 + 8 z^2 y^6\right),
		\label{scalarpot-smooth3}
	\end{equation}
	where $V_0 = \kappa^2 M_{*}^4$ and $M = \sqrt{M_{*} m_P}$. The potential is displayed in Figure \ref{fig:pot_plot} which shows two valleys of minima, approximated in the large $z$ limit as
	\begin{equation}
		y_{\pm} (z) = \pm\left( \sqrt{1 + 9 z^4} - 3 z^2\right)^{1/2} \simeq \pm \frac{1}{\sqrt{6} z}.
	\end{equation}
	This valley of local minimum is not flat and possess a slope that drives the inflaton towards the SUSY vacuum. Along the inflationary trajectory, SUSY is broken due to the presence of a non-zero vacuum energy density $V_0 = \kappa^2 M_{*}^4$, which in turn generates radiative corrections. These radiative corrections, however, are expected to have a negligible effect on the inflationary predictions, and therefore, we can safely ignore these contributions in our numerical calculations. Furthermore, along this inflationary trajectory ($y = y_+, z \gg 1$), the $G_{422}$ gauge symmetry breaks during inflation, and the potentially catastrophic magnetic monopoles resulting from the breaking of both $SU(4)_C$ and $SU(2)_R$ symmetries are effectively inflated away.
	
	Assuming gravity-mediated SUSY breaking \cite{Chamseddine:1982jx, Linde:1997sj}, where SUSY is broken in the hidden sector and is communicated gravitationally to the observable sector, the soft potential is \cite{Nilles:1983ge}
	\begin{equation}
		V_{\text{Soft}} = M_{\varphi_{i}}^2\lvert \varphi_{i} \rvert^2+m_{3/2}\big( \varphi_{i}W_{i}+\left(A-3\right)W + h.c.\big),
	\end{equation}
	where $\varphi_{i}$ is observable sector field, $W_{i}=\partial W/\partial \varphi_{i}$, $m_{3/2}$ is the gravitino mass and $A$ is the complex coefficient of the trilinear soft SUSY-breaking terms. The effective contributions of soft SUSY breaking terms during inflation can be written as,
	\begin{equation}
		V_{\text{Soft}} =  a m_{3/2} \kappa M_{*}^2 \lvert S \rvert + M_S^2 \lvert S \rvert^2=
		a m_{3/2} \kappa M_{*}^2 M z + M_S^2 M^2  z^2 ,
		\label{soft_terms}
	\end{equation}
	with
	\begin{equation*}
		a=2\vert A-2 \vert \cos \left(\arg S + \arg \vert A-2 \vert\right) ,
	\end{equation*}
	where $a$ and $M_S$ are the coefficients of soft SUSY breaking linear and mass terms for $S$, respectively. Here, we assume appropriate initial conditions for $\arg S$ so that $a$ remains constant during inflation \cite{REHMAN:2010191, urRehman:2006hu}. For the impact of $\arg S$ on the inflationary dynamics in standard hybrid inflation, see \cite{Buchmuller:2014epa}. 
	

	To describe physics near the Planck scale, it is crucial to incorporate supergravity (SUGRA) corrections, as they have an important effect on the global SUSY potential.
	
	The non-minimal K\"ahler potential may be expanded as
	\begin{align}\label{fullkahler} 
		K&= K_M + \kappa_S  \frac{|S|^4}{4m_P^2}+\kappa_{H^c}  \frac{|H^c|^4}{4m_P^2}+\kappa_{\bar{H}^c} \frac{|\bar{H}^c|^4}{4m_P^2}+\kappa_h  \frac{|h|^4}{4m_P^2}\nonumber \\ 
		&+\kappa_{S H^c}  \frac{|S|^2|H^c|^2}{m_P^2}+\kappa_{S\bar{H}^c}  \frac{|S|^2|\bar{H}^c|^2}{m_P^2}+\kappa_{Sh}  \frac{|S|^2|h|^2}{m_P^2}\nonumber \\
		&+\kappa_{H^c\bar{H}^c}  \frac{|H^c|^2|\bar{H}^c|^2}{m_P^2}+\kappa_{H^c h}  \frac{|H^c|^2|h|^2}{m_P^2}+\kappa_{\bar{H}^c h}  \frac{|\bar{H}^c|^2 |h|^2}{m_P^2}\nonumber\\
		&+\kappa_{SS} \frac{|S|^6}{6m_P^4}+... ~,
	\end{align}
	where $K_M$ is the minimal K\"ahler potential and is given as
	\begin{equation}\label{mkahler}
		K_M =  \lvert S\rvert^2+\lvert H^c \rvert^2+\lvert \bar{H}^c\rvert^2 +\lvert h\rvert^2.
	\end{equation}
	
	The $F$-term SUGRA scalar potential is given by 
	\begin{equation}
		V_{\text{SUGRA}}=e^{K/m_P^{2}}\left(
		K_{i\bar{j}}^{-1}D_{\varphi_{i}}WD_{\varphi^{*}_j}W^{*}-3 m_P^{-2}\left| W\right| ^{2}\right),
		\label{VF}
	\end{equation}
	with $\varphi_{i}$ being the bosonic components of the superfields $\varphi _{i}\in \{S, H^c,\bar{H}^c, h ,\cdots\}$, and we have defined
	\begin{equation}
		D_{\varphi_{i}}W \equiv \frac{\partial W}{\partial \varphi_{i}}+m_P^{-2}\frac{%
			\partial K}{\partial \varphi_{i}}W , \,\,\,
		K_{i\bar{j}} \equiv \frac{\partial ^{2}K}{\partial \varphi_{i}\partial \varphi_{j}^{*}},
	\end{equation}
	and $D_{\varphi_{i}^{*}}W^{*}=\left( D_{\varphi_{i}}W\right)^{*}.$
	The SUGRA scalar potential during inflation becomes
	\begin{align}
		V_{\text{SUGRA}} &= V_0 \left(1 - \frac{1}{54 z^4} + \left(\frac{8 + 3\, \kappa_S}{54 z^2} - \kappa_S\, z^2 \right)\left(\frac{M}{m_{p}}\right)^2 \right. \nonumber \\
		& \qquad \qquad \qquad \quad \left. +  \left(\frac{1 - 4 \kappa_S + 6 \gamma_S \, z^4}{12}\right) \left(\frac{M}{m_{p}}\right)^4 + \cdots\right)\, .
		\label{fullsugrapotential}
	\end{align}
	Putting soft SUSY breaking mass terms and SUGRA corrections together, we obtain the following form of inflationary potential,
	\begin{align}
		V &\simeq V_{\text{SUGRA}} + V_{\text{Soft}}  \nonumber\\
		&\simeq V_0 \left(1 - \frac{1}{54 z^4} + \left(\frac{8 + 3\, \kappa_S}{54 z^2} - \kappa_S\, z^2 \right)\left(\frac{M}{m_{p}}\right)^2 \right. \nonumber \\
		& \qquad \left. +  \left(\frac{1 - 4 \kappa_S + 6 \gamma_S \, z^4}{12}\right) \left(\frac{M}{m_{p}}\right)^4 + \frac{a\, m_{3/2} M }{M_{*}^2}z + \frac{M_S^2 M^2}{M_{*}^4}  z^2\right) ,
		\label{fullscalarpotential}
	\end{align}
	where $\gamma_S = 1 - \frac{7 \kappa_S}{2} + 2 \kappa_S^2 - 3 \kappa_{SS}$ and we have retained terms up to $\mathcal{O} ((\lvert S\rvert / m_P)^4)$ from SUGRA corrections. The dominant contribution to the potential arises solely from the terms involving higher powers of $S$ as all other fields ($\lvert H^c \rvert \sim (M/\lvert S \rvert) M$) are significantly suppressed in comparison ($\lvert S \rvert \gg M$).
	
	The inclusion of supergravity corrections often leads to the so called $\eta$ problem \cite{Linde:1997} by generating a large inflaton mass, which could spoil inflation. In the case of SUSY hybrid inflation with a minimal K\"ahler potential, as a direct consequence of $U(1)_R$ symmetry, a cancellation of the mass squared term through the interplay of the exponential factor and other part of the potential naturally circumvents this problem. On the other hand, for a non-minimal K\"ahler potential, this mass squared term appears with a coupling $\kappa_S$ which can be tuned ($\kappa_S \lesssim 0.01$) to ensure adequate flatness of the potential, required for successful inflation.
	
	\subsection{MSSM $\mu$-Term} 
	The MSSM $\mu$-term within the framework of the smooth hybrid inflation model can be derived as follows. The SUSY potential \eqref{scalarpot-smooth} in the $D$-flat direction, combined with the soft mass terms from Eq. \eqref{soft_terms}, is given by:  
	\begin{eqnarray}
		V_{\text{total}} &=& V_{\text{SUSY}} + V_{\text{Soft}} \nonumber \\
		&=& \kappa^2\, \Big| M_{*}^2 - \frac {(H^c)^4}{m_P^2} + \frac{\lambda}{\kappa} h^2 \Big|^2  + \frac{8 \, \kappa^2 \lvert S \rvert^2 \, \lvert H^c\rvert^6}{m_P^4} + 4 \lambda^2 h^2 \left| S \right|^2 \nonumber \\
		&+& \frac{a m_{3/2} \kappa M^4 \lvert S \rvert}{m_P^2} + M_S^2 \lvert S \rvert^2.
		\label{scalarpot-dflat}
	\end{eqnarray}
	 During inflation, the soft mass terms are suppressed, but they can induce a small shift in the VEV of $S$ from zero. Substituting the SUSY VEVs of $H^c$, $\bar{H}^c$, and $h$ from Eq. \eqref{gmin} into $V_{\text{total}}$ and aligning $S$ along the real axis via an appropriate $R$-transformation, the total potential simplifies to:
	\begin{eqnarray}
		V_{\text{total}} (S) &=& \frac{8  \kappa^2  S^2  M^6}{m_P^4} + \frac{a m_{3/2} \kappa M^4 S}{m_P^2} ,
	\end{eqnarray}
	where \(a = 1\) and \(M_S \ll M\) are assumed. Minimizing $V_{\text{total}}$ with respect to $S$ yields a non-zero VEV for $S$:  
	\begin{eqnarray}
		\frac{d}{dS} V_{\rm Total} (S) = 0 \quad \Rightarrow \quad \langle S \rangle \simeq -\frac{m_{3/2}}{16 \kappa} \left(\frac{m_P}{M}\right)^2 .
	\end{eqnarray}
	The $\mu$ term is then generated from the $\lambda S h^2$ term in Eq. \eqref{SP1}:
	\begin{equation}\label{mu_term_SHIM}
		\mu \simeq \lambda \langle S \rangle = -\frac{\lambda m_{3/2}}{16 \kappa} \left(\frac{m_P}{M}\right)^2 = -\frac{\lambda m_{3/2}}{16 \kappa} \left(\frac{M}{M_*}\right)^2.
	\end{equation}
	
	The stability of the inflationary trajectory with respect to fluctuations in the $h_u$ and $h_d$ fields can be evaluated by computing their mass spectrum during inflation. The mass-squared of these fields during inflation is given by
	\begin{equation}
		m_{h \pm}^2 = S^2 \lambda^2 \pm \kappa \lambda \left( M_{*}^2 - \frac{M^8}{36\, m_p^2 S^4} \right) \simeq S^2 \lambda^2 \pm \kappa \lambda M_{*}^2,
	\end{equation}  
	where the term $\kappa \lambda (M^4 / 6 m_p S^2)^2$ is highly suppressed and can be ignored. The mass-squared $m_{h -}^2 > 0 $ requires, 
	\begin{equation}\label{stability_con_SHI}
		\lambda > \frac{\kappa M_{*}^2}{S^2} .
	\end{equation}	
	
	In the smooth $\mu$-hybrid inflation model, the $H^c$-$\bar{H}^c$ system does not have a critical point, unlike the standard and shifted $\mu$-hybrid inflation scenarios, which feature two critical points and a hierarchy between $\kappa$ and $\lambda$ to ensure $H^c$-$\bar{H}^c$ stabilizes before $h_u$-$h_d$. Instead, the smooth model has a single critical point in the $h_u$-$h_d$ system, with $\lambda$ constrained to satisfy $S_{\text{end}} > S_c$.
	
	\subsection{Reheating and Non-Thermal Leptogenesis}
	
	After the end of inflation, the inflaton system, composed of two complex scalar fields: $S$ and $\theta=(\delta \phi + \delta \bar{\phi})/\sqrt{2}$ (where $\delta \phi = \langle H^c \rangle - M$ and $\delta \bar{\phi} = \langle \bar{H}^c \rangle - M$) with mass $m_I$, descends toward the SUSY minimum, experiences damped oscillations about it, and eventually undergoes decay, initiating the process referred to as 'reheating'. We consider the non-thermal leptogenesis scenario \cite{Senoguz:2004hq, lazarides:1991fx, lazarides:1997dx} to account for the observed baryon asymmetry. The superpotential term $\lambda S h^2$ induces an inflaton decay into a pair of higgsinos ($\widetilde h_u$, $\widetilde h_d$) and higgses ($ h_u$, $ h_d$), each with a decay width, $\Gamma_h$, given by \cite{Lazarides:1998qx},
	\begin{equation}\label{gamma}
		\Gamma_h =\Gamma(\theta \rightarrow \widetilde h_u\widetilde h_d) = \Gamma(S \rightarrow h_u h_d)=\frac{\lambda ^2 }{8 \pi }m_{I},
	\end{equation}
	where the inflaton mass $m_{I}$ is given by
	\begin{equation}
		m_{I} =2 \sqrt{2}\, \kappa \left(\frac{M_{*}^2}{M}\right) = 2 \sqrt{2}\, \kappa \left(\frac{M^3}{m_P^2}\right).
		\label{inf_mass}
	\end{equation}
	The right-handed neutrino mass receives contribution from the following superpotential term,
	\begin{equation}
	W \supset	\alpha^{ij}_1 F^{c}_i F^{c}_j \frac{\bar{H}^{c} \bar{H}^{c}}{m_P}   \supset M_{\nu_i}  \nu^{c}_i \nu^{c}_j,
	\end{equation}
	with a Majorana mass matrix $\alpha_{ij} (M^2/m_P)$ and eigenvalues
	\begin{equation}
		M_{\nu_i} =  \alpha_i  \frac{M^2}{m_P}.
	\end{equation}
	We assume hierarchical RHN Majorana masses with $M_{\nu_1} \ll M_{\nu_2}, M_{\nu_3}$ and $M_{\nu_1} > T_{\text{RH}}$.
	The inflaton decays into a pair of right-handed neutrinos ($\nu^c_i$) and sneutrinos ($\widetilde{\nu}^c_i$) with an equal decay width given as, 
	\begin{align}
		\Gamma_{\nu_i}& = \Gamma (\theta \rightarrow \nu^c_i \nu^c_i )= \Gamma(S \rightarrow \widetilde{\nu}^c_i \widetilde{\nu}^c_i ) \notag\\
		&=\dfrac{m_{I}}{8\pi}\left( \frac{M_{\nu_i}}{M} \right)^2\left(1-\dfrac{4M_{\nu_i}^{2}}{m_{I}^{2}}\right)^{1/2},
	\end{align}
	provided that only the lightest right-handed neutrino with mass $M_{\nu_1}$ satisfies the kinematic bound, $m_I > 2 M_{\nu_1}$. 
	
	With $H = 3\Gamma_I$, the reheat temperature $T_{\text{RH}}$ in expressed in terms of the total decay width of the inflaton, $\Gamma_I$, as
	\begin{equation}\label{tr}
		T_{\text{RH}}=\left(\dfrac{90}{\pi^{2}g_{*}}\right)^{1/4}\sqrt{\Gamma_I \, m_{P}},
	\end{equation}
	where $\Gamma_I = \Gamma_h + \Gamma_{\nu_1}$, and $g_{*}$ is the effective number of relativistic degrees of freedom at the temperature $T_{\text{RH}}$, which, for the MSSM spectrum, is $g_{*} = 228.75$. Assuming a standard thermal history, the number of e-folds, $N_{0}$, can be written in terms of the reheat temperature, $T_\text{RH}$, as \cite{Liddle:2003as},
	\begin{align}\label{efolds_tr}
		N_0 = 53 + \dfrac{1}{3}\ln\left[\dfrac{T_\text{RH}}{10^9 \text{ GeV}}\right]+\dfrac{2}{3}\ln\left[\dfrac{\sqrt{\kappa} M_{*}}{10^{15}\text{ GeV}}\right].
	\end{align}
	
	While $\Gamma_h$ is the dominant decay channel ($\Gamma_I \simeq \Gamma_h$), the $\Gamma_{\nu_1}$ channel is important for successful leptogenesis, which, through the sphaleron process \cite{Kuzmin:1985mm, Fukugita:1986hr, Khlebnikov:1988sr}, partially converts into the observed baryon asymmetry of the Universe (BAU). With $M_{\nu_1} \gg T_{\text{RH}}$, the washout factor of lepton asymmetry can be suppressed. 
	The observed baryon asymmetry is evaluated in terms of the lepton asymmetry factor, $\varepsilon_L$ \cite{mansoor:2020},
	\begin{align}\label{bphr}
		\frac{n_{B}}{n_{\gamma}}\simeq -1.84 \,\varepsilon_L  \frac{\Gamma_{\nu_1}}{\Gamma_I}\frac{T_{\text{RH}}}{m_I} \delta_{\text{eff}},
	\end{align} 
	where $|\delta_{\text{eff}}|\leq 1$ is the CP violating phase factor and the observed baryon-to-photon ratio, $n_\mathrm{B} /n_\gamma = (6.12 \pm 0.04) \times 10^{-10}$ \cite{ParticleDataGroup:2020ssz}. For hierarchical neutrino masses, the lepton-number asymmetry $\varepsilon_L$ is given by \cite{mansoor:2020, Luty:199245, covi:1996384},
	\begin{eqnarray}\label{lasymmetryfactor}
		\varepsilon_L  \simeq -\frac{3}{8\pi}  \frac{\sqrt{\Delta m_{31}^{2}} M_{\nu_1}}{\langle h_{u}\rangle^{2}}.
	\end{eqnarray} 
	Here, the atmospheric neutrino mass squared difference is $\Delta m_{31}^{2}\approx 2.6 \times 10^{-3}$ eV$^{2}$ and $\langle h_{u}\rangle \simeq 174$ GeV in the large $\tan\beta$ limit. For $|\delta_{\text{eff}}| = 1$ and using $\Gamma_h/\Gamma_{\nu_1} \simeq \lambda^2 \left(M/M_{\nu_1}\right)^2$, we obtain the right-handed neutrino mass in terms of the reheat temperature
	\begin{align}\label{lept}
		M_{\nu_1}  \simeq \left(\frac{n_{B}}{n_{\gamma}} \left(\frac{8 \pi}{5.52}\right) \frac{\lambda^2 \langle h_{u}\rangle^2}{\sqrt{\Delta m_{31}^{2}}}  \frac{ m_{I}M^2}{T_{\text{RH}}}\right)^{1/3}.
	\end{align} 
	For successful leptogenesis, the above equation must satisfy the kinametic bound $m_{I} \geq 2 M_{\nu_1}$. 
	\subsection{Inflationary Observables}
	
	The prediction of various inflationary parameters can now be estimated using standard slow-roll definitions described below. The leading order slow-roll parameters are defined as
	\begin{eqnarray}\label{eq:slow_roll_1}
		\epsilon (z) &=& \frac{1}{4}\left( \frac{m_P}{M}\right)^2
		\left( \frac{V' (z)}{V(z)}\right)^2, \quad
		\eta (z) = \frac{1}{2}\left( \frac{m_P}{M}\right)^2
		\left( \frac{V'' (z)}{V(z)} \right), \nonumber \\
		\xi^2 (z) &=& \frac{1}{4}\left( \frac{m_P}{M}\right)^4
		\left( \frac{V' (z) V''' (z)}{V^2 (z)}\right),
	\end{eqnarray}
	where the derivatives are with respect to $z = \lvert S \rvert / M$. In the slow-roll (leading order) approximation, the tensor-to-scalar ratio $r$, the scalar spectral index $n_s$, and the running of the scalar spectral index $dn_s / d \ln k$ are given by
	\begin{eqnarray}\label{tensor_to_scalar}
		r &\simeq& 16\,\epsilon (z),  \\\label{spectral_index}
		n_s &\simeq& 1 + 2 \,\eta (z) - 6\,\epsilon (z),  \\
		\frac{d n_s}{d\ln k} &\simeq& 16\,\epsilon (z)\,\eta (z)
		-24\,\epsilon^2 (z) - 2\,\xi^2 (z).
	\end{eqnarray}
	The amplitude of the curvature perturbation is given by
	\begin{equation}
		A_{s}(k_0) = \frac{1}{24\,\pi^2}
		\left. \left( \frac{V/m_P^4}{\epsilon}\right)\right|_{z = z_0}.
		\label{perturb}
	\end{equation}
	where $A_{s} = 2.1 \times 10^{-9}$ is the Planck normalization at $k_0 = 0.05\ \rm{Mpc}^{-1}$ \cite{Planck:2018jri}.
	The last $N_0$ number of $e$-folds before the end of inflation is,
	\begin{equation}
		N_0 = 2\left( \frac{M}{m_P}\right) ^{2}\int_{z_e}^{z_{0}}\left( \frac{V}{%
			V'}\right) dz,
		\label{efolds}
	\end{equation}
	where $z_0$ is the field value at the pivot scale $k_0$, and $z_e$ is the field value at the end of inflation, defined by $|\eta(z_e)| = 1$.
	
	\subsection{Inflationary Predictions and Comparison with Observations}
	
	The constraints on the scalar spectral index $n_s$ and the tensor-to-scalar ratio $r$ in the $\Lambda$CDM $+ r$ model from the Planck 2018 CMB power spectra, in combination with CMB lensing
	reconstruction and BICEP2/Keck Array (BK15) data at the 95 \% confidence level, are \cite{Planck:2018jri}:
	\twotwosig[5.0cm]{n_s &= 0.9651 \pm 0.0041 ,}{r &<
		0.061,}{\textit{Planck} TT,TE,EE +lowE+lensing+BK15.  \label{PLANCK_POL_BK15}}
	
	It can be easily verified that, in the case of the minimal K\"ahler potential (with $\kappa_{S} = \kappa_{SS} =0$), SUGRA corrections dominate the global SUSY potential. This leads to the prediction of $n_s$ and $r$ inconsistent with the above bounds. In simpler terms, the SUGRA corrections necessitate trans-Planckian field values to achieve $n_s$ and $r$ within experimental bounds, thereby invalidating the SUGRA expansion itself. Consequently, the minimal case ($\kappa_{S} = \kappa_{SS} =0$) is inconsistent with the current observations.
	
\begin{figure}[t]
			\centering \includegraphics[width=\textwidth]{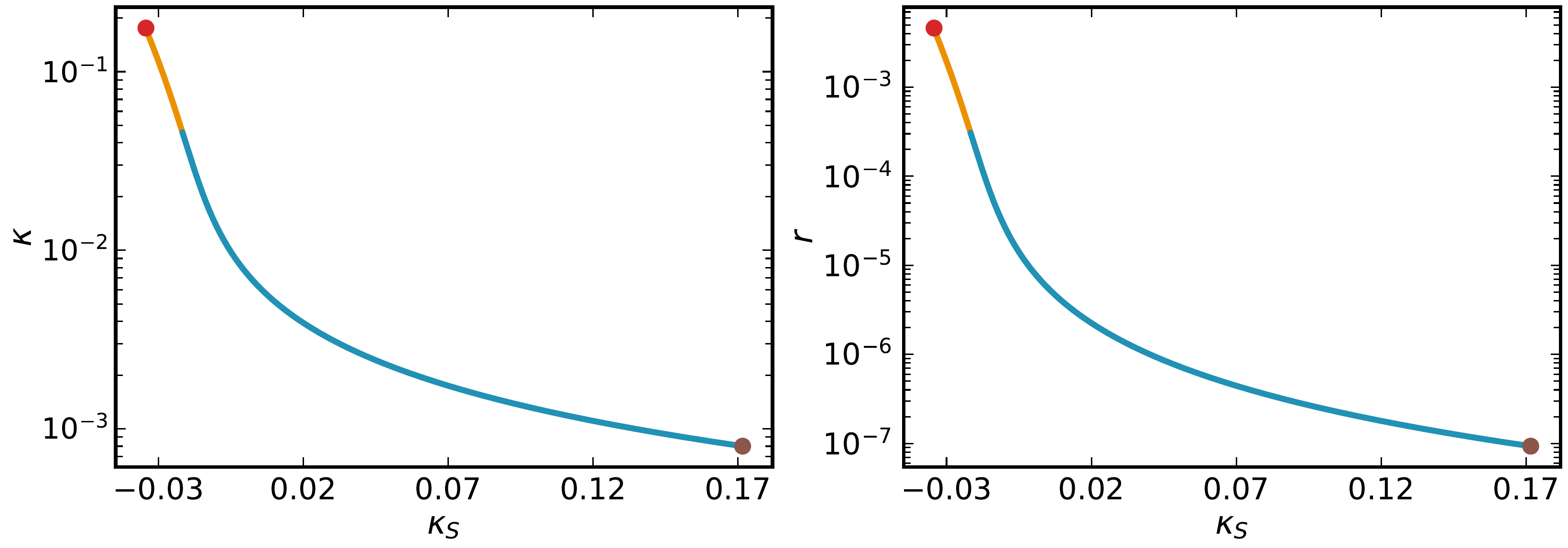}
			\caption{Variation of dimensionless parameter $\kappa$ (left) and tensor-to-scalar ratio $r$ (right) with the non-minimal coupling $\kappa_S$. The yellow and blue curves represent field values $\lvert S_0 \rvert \gtrsim 0.5\, m_P$ and $\lvert S_0 \rvert \lesssim 0.5\, m_P$, respectively. The red and brown circles correspond to $\lvert S_0 \rvert  = m_P$ and $\kappa_{SS} = -1$ constraints, respectively.}
			\label{fig:nonminimal_kahler_1}
		\end{figure}

\begin{figure}[t]
			\centering \includegraphics[width=\textwidth]{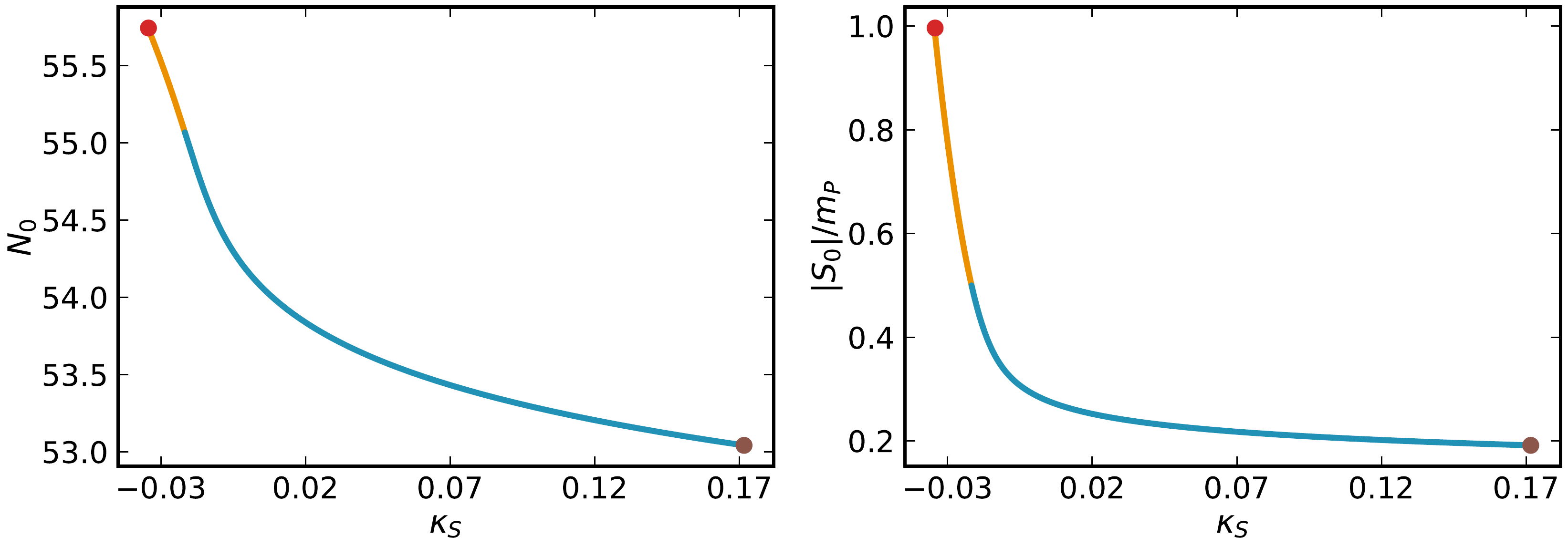}
			\caption{Variation of number of $e$-folds $N_0$ (left) and $\lvert S_0 \rvert  / m_P$ (right) with the non-minimal coupling $\kappa_S$. The yellow and blue curves represent field values $\lvert S_0 \rvert \gtrsim 0.5\, m_P$ and $\lvert S_0 \rvert \lesssim 0.5\, m_P$, respectively. The red and brown circles correspond to $\lvert S_0 \rvert  = m_P$ and $\kappa_{SS} = -1$ constraints, respectively.}
			\label{fig:nonminimal_kahler_2}
		\end{figure}

\begin{figure}[t]
			\centering \includegraphics[width=\textwidth]{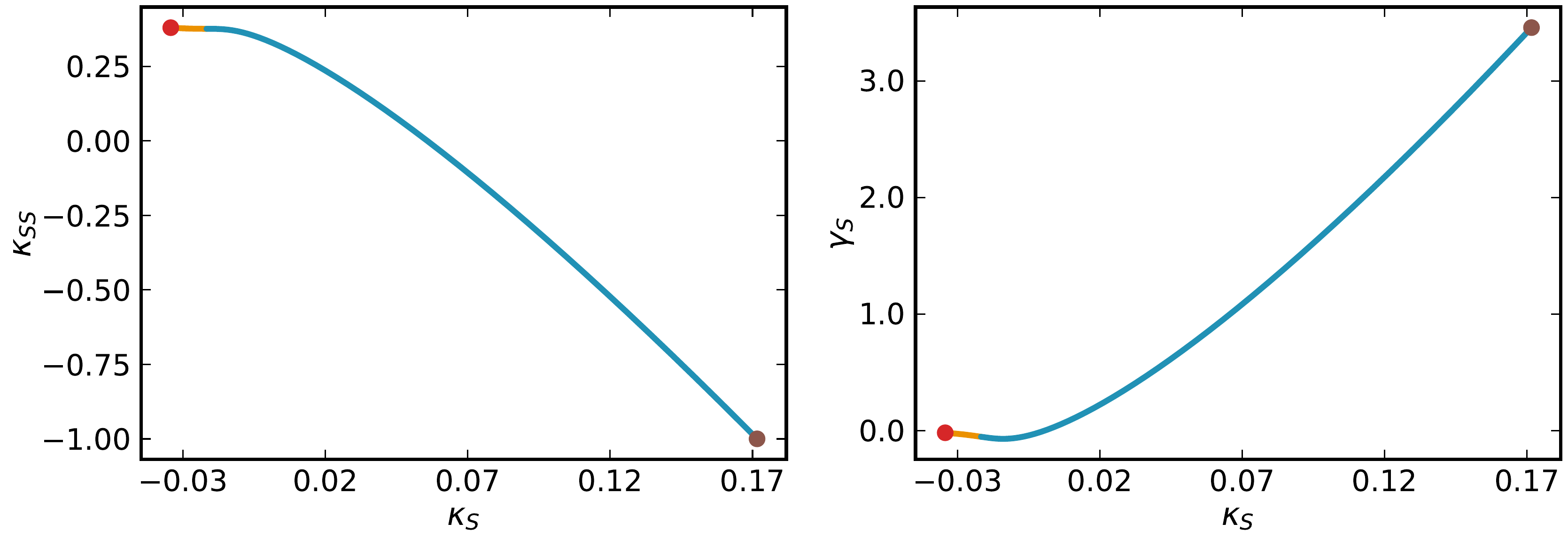}
			\caption{Variation of non-minimal coupling $\kappa_{SS}$ (left) and quartic coupling $\gamma_S$ (right) with the non-minimal coupling $\kappa_S$. The yellow and blue curves represent field values $\lvert S_0 \rvert \gtrsim 0.5\, m_P$ and $\lvert S_0 \rvert \lesssim 0.5\, m_P$, respectively. The red and brown circles correspond to $\lvert S_0 \rvert  = m_P$ and $\kappa_{SS} = -1$ constraints, respectively.}
			\label{fig:nonminimal_kahler_3}
		\end{figure}

\begin{figure}[t]
			\centering \includegraphics[width=\textwidth]{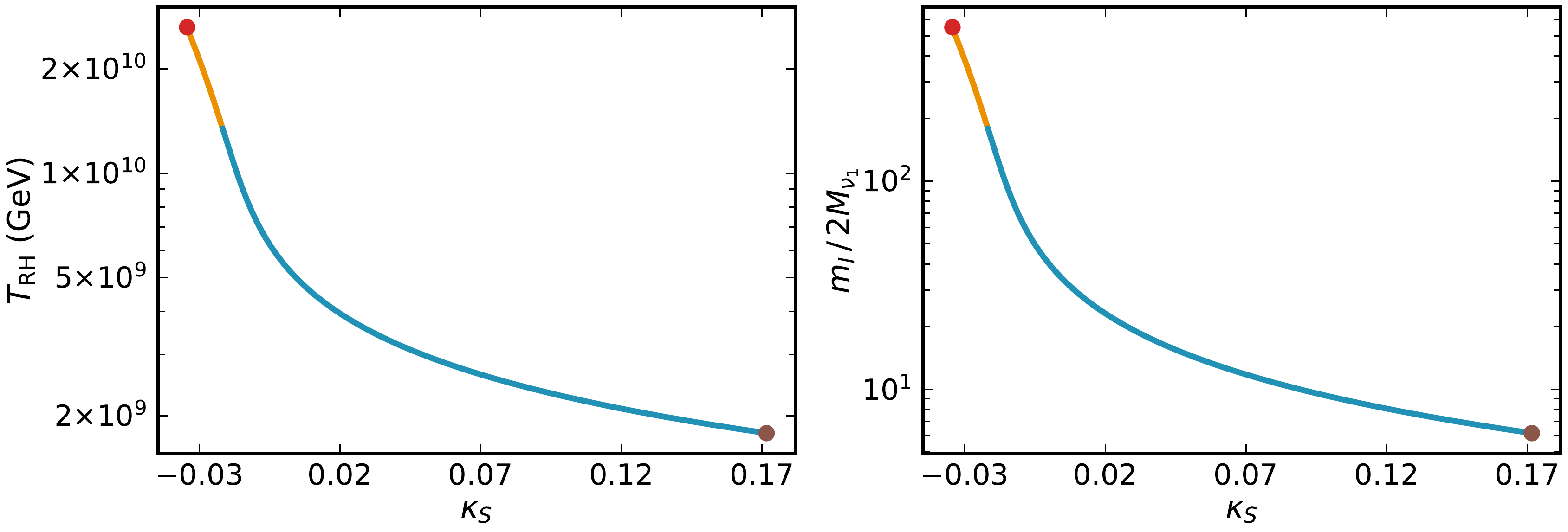}
			\caption{Variation of the reheat temperature $T_\text{RH}$ (left) and the inflaton mass to RHN mass ratio $m_{I}/ 2 M_{\nu_1}$ (right) with the non-minimal coupling $\kappa_S$. The yellow and blue curves represent field values $\lvert S_0 \rvert \gtrsim 0.5\, m_P$ and $\lvert S_0 \rvert \lesssim 0.5\, m_P$, respectively. The red and brown circles correspond to $\lvert S_0 \rvert  = m_P$ and $\kappa_{SS} = -1$ constraints, respectively.}
			\label{fig:nonminimal_kahler_4}
		\end{figure}

\begin{figure}[t]
			\centering \includegraphics[width=\textwidth]{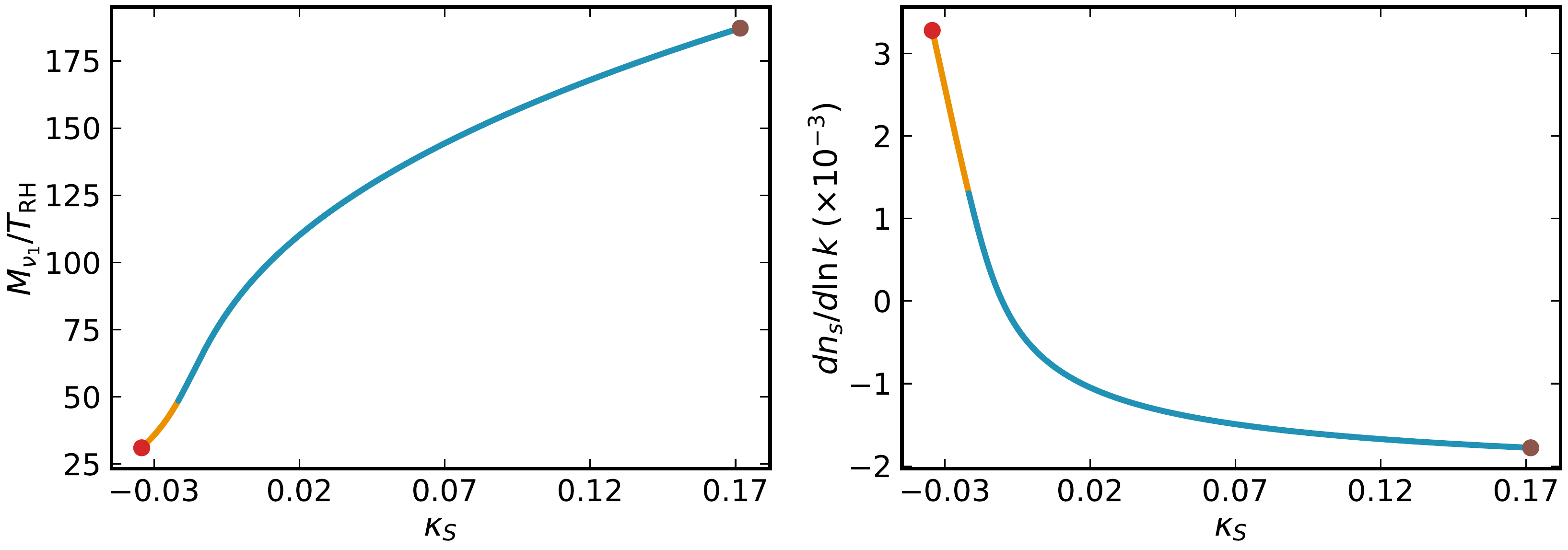}
			\caption{Variation of the ratio $M_{\nu_1} / T_\text{RH}$ (left) and the running of the scalar spectral index $dn_s/d\ln k$ (right) with the non-minimal coupling $\kappa_S$. The yellow and blue curves represent field values $\lvert S_0 \rvert \gtrsim 0.5\, m_P$ and $\lvert S_0 \rvert \lesssim 0.5\, m_P$, respectively. The red and brown circles correspond to $\lvert S_0 \rvert  = m_P$ and $\kappa_{SS} = -1$ constraints, respectively.}
			\label{fig:nonminimal_kahler_5}
		\end{figure}

	The results of our numerical calculations for the inflationary potential in Eq. \eqref{fullscalarpotential} with a non-minimal K\"ahler potential are displayed in Figs. \ref{fig:nonminimal_kahler_1} - \ref{fig:nonminimal_kahler_5}. To obtain these results, a second-order approximation has been employed for the slow-roll parameters, and the scalar spectral index $n_s$ has been fixed at the central value ($\sim 0.9651$) of Planck's 2018 data bounds. Additionally, the soft SUSY masses have been set at $m_{3/2} \simeq M_S \simeq 100$ TeV, with $a = 1$. Note that for $m_{3/2} \simeq 100$ TeV, there is no upper bound on the reheat temperature $T_{\text{RH}}$ \cite{ellis:1984kn, Linde:1984yk}.  We further set the mass parameter $M_{*} \simeq 2 \times 10^{16}$ GeV, which in turn determines $M = \sqrt{M_{*} m_p} \simeq 2.2 \times 10^{17}$ GeV. The right-hand side of the inequality \eqref{stability_con_SHI} varies between $\simeq 1.2 \times 10^{-5} - 1.5 \times 10^{-6}$ for the range of parameters obtained in our numerical calculations. Therefore, we set $\lambda \sim 2 \times 10^{-5}$ in our analysis. This choice ensures that both the kinematic bound ($m_{I} \geq 2 M_{\nu_1}$) and the out-of-equilibrium condition for leptogenesis ($M_{\nu_1} > T_{\text{RH}}$) are easily met.
	
	The SUGRA corrections, parameterized by $\kappa_{S}$ and $\kappa_{SS}$, dominate the global SUSY potential, while the soft mass terms with $m_{3/2} \simeq M_S \simeq 100$ TeV are highly suppressed. To keep the SUGRA expansion under control, we impose $\lvert S_0 \rvert \leq m_P$. Additionally, we restrict the non-minimal couplings $\lvert \kappa_{S} \rvert \leq 1$ and $\lvert \kappa_{SS} \rvert \leq 1$. These constraints are depicted in Figs. \ref{fig:nonminimal_kahler_1} - \ref{fig:nonminimal_kahler_5} by the red ($\lvert S_0 \rvert = m_P$) and brown  ($\kappa_{SS} = -1$) circles. The yellow region, with $\lvert S_0 \rvert \gtrsim 0.5\, m_P$, represents the ultraviolet sensitivity region where higher-order Planck-suppressed terms in the SUGRA expansion become important. On the other hand, the blue region represents $\lvert S_0 \rvert \lesssim 0.5\, m_P$, where a natural suppression of higher-order terms is achieved along with the boundedness of the potential. This issue arises due to the quartic coupling being $\gamma_S < 0$. It is evident that both the quadratic ($\kappa_S$) and quartic ($\gamma_{S}$) couplings play a vital role in bringing the scalar spectral index $n_s$ within the Planck bounds, accompanied by a large value of the tensor-to-scalar ratio $r$.
		
		The behavior of the tensor-to-scalar ratio $r$ and the dimensionless parameter $\kappa$, as displayed in Fig. \ref{fig:nonminimal_kahler_1}, can be understood through the following explicit relationship between $r$, $\kappa$ and $M$
		\begin{equation}
			r \simeq \left(\frac{2 \kappa^2}{3 \pi^2 A_s(k_0)}\right) \left(\frac{M}{m_P}\right)^8 \simeq \kappa^2 \left(\frac{M}{3.23 \times 10^{16} \text{ GeV}}\right)^4 \left(\frac{M}{m_P}\right)^4.
		\end{equation}
		Since $M$ is fixed, larger values of $r$ are expected for large values of $\kappa$. It can readily be checked that the largest value of $r$ ($\sim 4 \times 10^{-3}$) obtained in our numerical results occurs for $\kappa \simeq 0.17$. The behavior of the last number of $e$-folds $N_0$ and $\lvert S_0 \rvert/m_P$ with respect to $\kappa_{S}$ is presented in Fig. \ref{fig:nonminimal_kahler_2}, while Fig. \ref{fig:nonminimal_kahler_3} depicts the behavior of $\kappa_{SS}$ and $\gamma_{S}$ with respect to $\kappa_{S}$. The number of $e$-folds vary within the range $53 \lesssim N_0 \lesssim 56$. It can be seen that the large $r$ values are obtained with non-minimal couplings $\kappa_S < 0$, $\kappa_{SS} > 0$ and the quartic coupling $\gamma_S <0 $. Also, the large $r$ values occur for tiny values of the quartic coupling $\gamma_S\, (\sim-0.02)$. To facilitate this discussion further, we provide an analytical estimate of the couplings $\kappa_{S}$ and $\gamma_{S}$ in the large $r$ region, where the contribution from the global SUSY part of the potential is negligible, and $\lvert S_0 \rvert$ approaches $m_P$. Using Eqs. \eqref{tensor_to_scalar}, \eqref{spectral_index} and \eqref{efolds} with $\lvert S_0 \rvert  \simeq m_P$, we obtain the following approximate expressions for number of $e$-folds $N_0$, scalar spectral index $n_s$ and tensor to scalar ratio $r$  
		\begin{align} \label{analytic_efolds}
			N_0 &\simeq \frac{1}{6\,\kappa_S} \log \left( - \left( \frac{2 \times 10^{17} \text{ GeV}}{m_P} \right)^2
			\frac{\left( \gamma_S -\kappa_S \right)^4}{\kappa_S^3} \right),\\
		 \label{analytic_spectral_index}
			n_s &\simeq 1 + \frac{10}{27 z_0^6} \left(\frac{m_P}{M}\right)^2 - 2 \kappa_S + 6 \gamma_S z_0^2 \left(\frac{M}{m_P}\right)^2 \simeq 1 - 2 \kappa_S + 6 \gamma_S, \\
			\label{analytic_tensor_modes}
			r &\simeq 16 \left(\frac{1}{27 z_0^5} \left(\frac{m_P}{M}\right)^3 - \kappa_S z_0 \left(\frac{M}{m_P}\right) + \gamma_S z_0^3 \left(\frac{M}{m_P}\right)^3 \right)^2 \simeq 16 \left(\gamma_S - \kappa_S\right)^2.
		\end{align}
		Solving Eqs. \eqref{analytic_efolds} and \eqref{analytic_spectral_index} simultaneously for $N_0 \simeq 55$ and $n_s \simeq 0.9651$, we obtain $\kappa_S \sim -0.034$ and $\gamma_S \sim -0.017$. Using these values in Eq. \eqref{analytic_tensor_modes}, we obtain $r \sim 0.004$ which is a reasonably good estimate of the otherwise more accurately calculated numerical results displayed in Figs. \ref{fig:nonminimal_kahler_1} - \ref{fig:nonminimal_kahler_5}. Note that the large tensor modes require values of $M$ greater than the GUT scale $M_{\text{GUT}} \simeq 2 \times 10^{16}$ GeV. Figure \ref{fig:nonminimal_kahler_4} illustrates the variation of the reheat temperature, $T_{\text{RH}}$, and the ratio of the inflaton mass to the right-handed neutrino (RHN) mass, $m_I / 2 M_{\nu_1}$. Meanwhile, Figure \ref{fig:nonminimal_kahler_5} shows the variation of the ratio $M_{\nu_1} / T_{\text{RH}}$ and the running of the scalar spectral index, $dn_s / d\ln k$, with respect to $\kappa_S$. The reheat temperature varies in the range $1.8 \times 10^{9} \text{ GeV} \lesssim T_{\text{RH}} \lesssim 2.6 \times 10^{10} \text{ GeV}$, with the kinematic bound $m_I \geq 2 M_{\nu_1}$ satisfied across the entire range of $\kappa_S$. Additionally, since $T_{\text{RH}} < M_{\nu_1}$, the out-of-equilibrium condition for successful leptogenesis is easily met. The running of the scalar spectral index ranges from $-0.0018 \lesssim dn_s / d\ln k \lesssim 0.0033$, consistent with the $\Lambda$CDM$+r$ model assumptions. The MSSM $\mu$-term of $\mathcal{O}(1)$ TeV can be generated from Eq. \eqref{mu_term_SHIM} with $\lambda \sim 2 \times 10^{-5}$, $\kappa \sim 0.015$, and a gravitino mass $m_{3/2} \simeq 100$ TeV.
		
		In summary, for non-minimal couplings $-0.034 \lesssim \kappa_S \lesssim 0.17$ and $ -1 \lesssim \kappa_{SS} \lesssim 0.38 $, with the scalar spectral index fixed at the central value ($n_s \simeq 0.9651$) of Planck 2018 data bounds and $M_{*}$ fixed at $2 \times 10^{16}$ GeV, we obtain a tensor-to-scalar ratio $9.3 \times 10^{-8} \lesssim r \lesssim 4.6 \times 10^{-3}$. The inflaton mass ranges from $4.1 \times 10^{12}$ GeV to $9.0 \times 10^{14}$ GeV, and the RHN mass ranges from $3.3 \times 10^{11}$ GeV to $8.2 \times 10^{11}$ GeV, with $\alpha_1$ in the range $1.7 \times 10^{-5} \lesssim \alpha_1 \lesssim 4.1 \times 10^{-5}$. The predictions of the model are thus perfectly compatible with current observations.
				
		\section{Non-Minimal Higgs Inflation} \label{sec4}
		
		We utilize the same superpotential $W$ as given in Eq. \eqref{SP1}. In non-minimal Higgs inflation, the conjugate Higgs fields ($H^c, \bar{H}^c$) assume the role of the inflaton. To implement non-minimal Higgs inflation in the context of $G_{422} \equiv SU(4)_C \times SU(2)_L \times SU(2)_R$ model with a global $U(1)_R$ and a $Z_2$ symmetry, we adopt the following form of the K\"ahler potential\footnote{It has recently been shown in \cite{mansoor:2024ni} that incorporating a sextic term ($|S|^6/m_P^6$) in the K\"ahler potential plays a crucial role in the generation of primordial black holes (PBH). In another study \cite{mansoor:2023nm}, a cubic term of the form $S^3/m_P^3$ is shown to play a similar role in generating PBHs, where $U(1)_R$ symmetry is assumed to be broken at the non-renormalizable level.} \cite{mansoor:201928}
		\begin{equation}\label{nmhi_kahlar}
			\begin{split}
				K=&-3m_{P}^{2}\log\left(1-\frac{(|S|^{2}+|H^c|^{2}+ |\bar{H}^c|^{2})}{3m_{P}^{2}}+\frac{\delta}{2m_{P}^{4}} \left( (H^c \bar{H}^c)^{2} + h.c \right)   \right. \\
				& + \left.  \frac{\gamma}{3m_{P}^{4}} \left( |S|^{4} + |S|^{2} |H^c|^{2} + |S|^{2} |\bar{H}^c|^{2} + |H^c|^{4}+ |\bar{H}^c|^{4} \right) \right),
			\end{split}
		\end{equation}
		where $\delta$ and $\gamma$ are dimensionless couplings that vanish in the exact no-scale limit and we have assumed the stabilization of the modulus fields \cite{Ellis:2013xoa, Ellis:2013oct, Cicoli:2013oct29}. The term with coupling $\delta$ plays a pivotal role in non-minimal Higgs inflation, while the higher-order term with the coupling $\gamma$ is required to stabilize $S$ during inflation \cite{Lee:2010hj}. It is worth noting that $S$ can also be stabilized without the inclusion of quartic terms, as shown in \cite{Linde:2024rs, kpallis_NToumbas:2016}.
		
		The scalar potential in the Einstein' frame is given by, 
		\begin{equation}
			V_E = e^{G/m_P^2}\left( G_i (G^{-1})_j^i  G^j- 3 m_P^4\right)+\frac{1}{2} g_a^2 G^i (T_a)_i^j z_j (\text{Re} f^{-1}_{ab}) G^k (T_b)_k^l z_l,
		\end{equation}
		with
		\begin{equation}
			G = K + m_P^2 \log\left(\frac{|W|^2}{m_P^6}\right), \quad  G^i = m_P \frac{\partial G}{\partial z_i} ,\quad G_i = m_P \frac{\partial G}{\partial z_i^*}, \quad G^i_j = \frac{\partial G}{\partial z_i \partial z_j^*},
		\end{equation}
		where $z_{i}\in\{ S, H^c,\bar{H}^c\}$, $f_{ab}$ represents the gauge kinetic function, and $T_a$ denotes the generators of the gauge group. Here, we have used the same notation for both the superfields and their corresponding scalar components. The complex Higgs fields can be expressed in terms of real scalar fields as follows:
		\begin{eqnarray}\label{real_scalar}
			H^c = \frac{w}{\sqrt{2}} \ e^{i \theta} \cos \phi, \qquad \bar{H}^c  = \frac{w}{\sqrt{2}} \ e^{i \bar \theta} \sin\phi.
		\end{eqnarray}
		In the $D$-flat direction, the phases can be stabilized at $\theta=\bar \theta=0$, $\phi = \pi/4$, leading to:
		\begin{equation} 
			H^c = \bar{H}^c = \frac{w}{2},
		\end{equation} 		
		where $w$ represents the real scalar field in the Jordan frame. The scalar potential in the Einstein frame then takes the form:
		\begin{eqnarray}\label{scalar_pot_nmhi}
			V_{E} &=& \frac{\kappa^2 M_{*}^{4}\left(1-\left(\frac{w}{2M}\right)^{4}\right)^{2} }{\left( 1 - 2 \gamma \left(\frac{w}{2m_{P}}\right)^{2} \right) \left(1-\frac{2}{3}\left(\frac{w}{2m_{P}}\right)^{2}+\frac{1}{3} \left(3\delta + 2 \gamma\right) \left(\frac{w}{2m_P}\right)^{4}\right)^{2}} \nonumber \\ 
			&=& \frac{\kappa^2 M_{*}^{4}}{\Omega^2}\left(1-\left(\frac{w}{2M}\right)^{4}\right)^{2} \left( 1 - 2 \gamma \left(\frac{w}{2m_{P}}\right)^{2} \right)^{-1},
		\end{eqnarray}
		where, $M = \sqrt{M_{*} m_P}$ and the conformal scaling factor, $\Omega = e^{-K/3 m_P^2}$, which relates the Einstein and Jordan frames, is given by
		\begin{equation}
			g^J_{\mu\nu} = \Omega g^E_{\mu\nu} = \left(1-\frac{2}{3}\left(\frac{w}{2m_{P}}\right)^{2}+\frac{1}{3} \left(3\delta + 2 \gamma\right) \left(\frac{w}{2m_P}\right)^{4}\right) g^E_{\mu\nu}.
		\end{equation}
		The action in the Einstein' frame is
		\begin{equation}\label{Jordan}
			S_E = \int d^4 x \sqrt{-g_E}\left[\dfrac{1}{2} m_P^2 \mathcal{R}(g_E) - \frac{1}{2} g_E^{\mu \nu} \partial_\mu \widehat{w}\, \partial_\nu \widehat{w} - V_E (\widehat{w}(w))\right],
		\end{equation}
		where $\widehat{w}$ is the canonically normalized inflaton field in the Einstein frame, defined as
		\begin{equation}
			\frac{d\widehat{w}}{dw} \equiv J (w) = \sqrt{\frac{1}{\Omega(w)}+\frac{3}{2}m_{P}^{2}\left(\frac{\Omega^{\prime}(w)}{\Omega (w)}\right)^{2}}.
		\end{equation}
		
		\subsection{MSSM $\mu$-Term}
		
		Following \cite{kpallis:2018}, we derive the MSSM $\mu$ term within the framework of the non-minimal Higgs inflation model. The scalar potential in the SUSY limit is expressed as \cite{spmartin:2010}:
		\begin{equation}\label{potential_susy_limit}
			V_{\rm SUSY}= \widetilde K^{\alpha \bar{\beta}}
			W_{\alpha} W^*_{\bar{\beta}}+\frac{g^2}2 \mbox{$\sum_{a}$}
			{D}_{a} {D}_{a},
		\end{equation}
		where the superpotential $W$ is provided in Eq. \eqref{SP1}, and $\widetilde{K}$ represents the $m_P \rightarrow \infty$ limit of the K\"ahler potential in Eq. \eqref{nmhi_kahlar}, given by:
		\begin{eqnarray}\label{kahlar_limit}
			\widetilde K &=& |S|^{2}+|H^c|^{2} +  |\bar{H}^c|^{2} - \frac{3\delta}{2m_{P}^{2}} \left( (H^c \bar{H}^c)^{2} + h.c \right) \nonumber   \\
			& -& \frac{\gamma}{m_{P}^{2}} \left( |S|^{4} + |S|^{2} |H^c|^{2} + |S|^{2} |\bar{H}^c|^{2} + |H^c|^{4}+ |\bar{H}^c|^{4} \right) + \cdots.
		\end{eqnarray}
		Using Eqs. \eqref{kahlar_limit} and \eqref{SP1} into Eq. \eqref{potential_susy_limit}, the SUSY potential in the $D$-flat direction, after substituting the VEVs from Eq. \eqref{gmin}, becomes
		\begin{eqnarray}
			V_{\rm SUSY} = 8 \kappa^2 S^2 M^2  \left(\frac{ M}{m_P}\right)^4 \left( 1 - 5\delta \left(\frac{M}{m_P}\right)^4\right)^{-1} .
		\end{eqnarray}
		The soft SUSY breaking terms, while negligible during inflation, can cause a slight shift in the VEV of $S$ from zero. Including the soft SUSY breaking terms from Eq. \eqref{soft_terms}, the total SUSY potential can be written as
		\begin{eqnarray}
			V_{\rm Total} (S) &=& V_{\rm SUSY} (S) + V_{\rm Soft} (S), \nonumber \\
			&=& 8 \kappa^2 S^2 M^2  \left(\frac{M}{m_P}\right)^4 \left( 1 - 5\delta \left(\frac{M}{m_P}\right)^4\right)^{-1} + \kappa m_{3/2} S M^2 \left(\frac{M}{m_P}\right)^2,
		\end{eqnarray}
		where we have used the fact that $M_S \ll M$ and have set $a = 1$. Minimizing the total potential with respect to $S$ yields the following non-zero VEV for $S$,
		\begin{eqnarray}
			\frac{d}{dS} V_{\rm Total} (S) = 0 \quad \Rightarrow \quad \langle S \rangle \simeq \frac{m_{3/2}}{16 \kappa} \left( 5\delta \left(\frac{M}{m_P}\right)^2 - \left(\frac{m_P}{M}\right)^2 \right).
		\end{eqnarray}
		The $\mu$ term is then generated from the $\lambda S h^2$ term in Eq. \eqref{SP1}:
		\begin{equation}\label{mu_term_NMHI}
			\mu \simeq \lambda \langle S \rangle = \frac{\lambda m_{3/2}}{16 \kappa} \left( 5\delta \left(\frac{M}{m_P}\right)^2 - \left(\frac{m_P}{M}\right)^2 \right).
		\end{equation}
		Note that in the limit $\delta \rightarrow 0$, the $\mu$ term above simplifies to the same expression derived in the smooth hybrid inflation model, as given in Eq. \eqref{mu_term_SHIM}.
		
		The stability of the inflationary trajectory with respect to fluctuations in the $h_u$ and $h_d$ fields can be evaluated by computing their mass spectrum during inflation. The mass-squared of these fields during inflation is found to be:
		\begin{equation}
			m_{h \pm}^2 = 4 H_I^2 \left( \frac{1}{\Omega(w)} \pm \frac{12 m_p^4 \lambda}{\kappa w^4} \right),
		\end{equation}  
		where $H_I^2 \simeq \frac{V_E}{3 m_P^2}$. The condition $m_{h -}^2 > 0$ requires: 
		\begin{equation}\label{stability_con_NMHI}
			\lambda < \frac{\kappa w^4}{12\, \Omega\,  m_P^4} \simeq \frac{4 \kappa}{3 \delta}.
		\end{equation}	
		
		\subsection{Reheating and Non-Thermal Leptogenesis}
		
		After the end of inflation, the inflaton continues to roll towards the SUSY vacuum, eventually settling into a phase of damped oscillations. Subsequently, the inflaton decays, initiating the reheating of the universe. The decay widths of the inflaton decay into a pair of higgsinos ($\widetilde h_u$, $\widetilde h_d$) and a pair of right-handed neutrinos ($\nu^c_i$) \cite{C.Pallis_2011} are respectively given by 
		\begin{eqnarray}\label{gamma_h}
			\Gamma_h &=& \frac{\lambda ^2 }{8 \pi \Omega_0^2}m_{I}, \\ \label{gamma_N}
			\Gamma_{\nu_i} &=& \dfrac{m_I}{64\pi}\left( \frac{M_{\nu_i}}{M} \frac{\Omega_0^{3/2}}{J_0} \right)^2 \left( 1 + \left( \frac{M}{m_P} \right)^2 - \frac{1}{2} \left(3\delta + 2\gamma\right)  \left( \frac{M}{m_P} \right)^4 \right)^2 \left(1-\dfrac{4M_{\nu_i}^{2}}{m_I^{2}}\right)^{1/2},
		\end{eqnarray}
		where the inflaton mass $m_{I}$ and the RHN mass $M_{\nu_i}$ are given as
		\begin{equation}
			m_{I} = \frac{4\, \kappa}{J_0 \Omega_0}\left(\frac{M_{*}^2}{M}\right) = \frac{4\, \kappa}{J_0 \Omega_0}\left(\frac{M^3}{m_P^2}\right), \qquad M_{\nu_i} = \alpha_i \frac{M^2}{m_P \sqrt{\Omega_0}} ,
			\label{inf_mass}
		\end{equation}
		with
		\begin{eqnarray}
				 \Omega_0 &=& \Omega (\langle w \rangle = 2 M) =  1 - \frac{2}{3} \left(\frac{M}{m_P} \right)^2 + \frac{1}{3} \left(3\delta + 2 \gamma\right) \left(\frac{M}{m_P} \right)^4, \\ 
				 J_0 &=& J (\langle w \rangle = 2 M).
		\end{eqnarray}
		In the supergravity framework, the inflaton can spontaneously decay into MSSM particles \cite{Endo:2006qk, Endo:2007sz}. In the current SUSY GUT model, the Yukawa interaction for the third generation, $y_{3 3}^{(u,\nu)} Q_3 L_3 H_u$, gives rise to the following 3-body decay width \cite{C.Pallis_2011},
		\begin{equation}
			\Gamma_{y_{t}}=\frac{7}{256 \pi^3} \left( \frac{y_{t}  \, \Omega_0^{3/2}}{J_0} \right)^2 \left( 1 - \left(3 \delta + 2 \gamma\right) \left( \frac{M}{m_P} \right)^2 \right)^2 
			\left( \frac{M}{m_P} \right)^2 \left( \frac{m_I}{m_P} \right)^2 m_I,
		\end{equation}
		where, $y_t=y_{3 3}^{(u,\nu)}$ corresponds to the top Yukawa coupling. The reheating temperature $T_{\text{RH}}$ is expressed in terms of the inflaton decay width in Eq. \eqref{tr}, where the total  decay width of the inflaton is now $\Gamma_I = \Gamma_{\nu_i} + \Gamma_h + \Gamma_{y_{t}}$. Using Eqs. \eqref{bphr}, \eqref{lasymmetryfactor}, \eqref{gamma_N}, and assuming hierarchical RHN Majorana masses with $M_{\nu_1} \ll M_{\nu_2}, M_{\nu_3}$ and $M_{\nu_1} > T_{\text{RH}}$, we obtain the right-handed neutrino mass in terms of the reheat temperature
		\begin{equation}\label{RHN_mass_leptogen}
			M_{\nu_1}  \simeq \left(\frac{n_{B}}{n_{\gamma}} \left(\frac{512 \pi^2}{5.52}\right) \frac{\left(\Gamma_h + \Gamma_{y_{t}}\right) \langle h_{u}\rangle^2}{\sqrt{\Delta m_{31}^{2}} T_{\text{RH}}}  \left(\frac{J_0 M}{\Omega_0^{3/2}}\right)^2 \left(1 + \left( \frac{M}{m_P} \right)^2 - \frac{1}{2} \left(3\delta + 2 \gamma\right) \left( \frac{M}{m_P} \right)^4\right)^{-2}\right)^{1/3},
		\end{equation}
		where $\Gamma_I \simeq \Gamma_h + \Gamma_{y_{t}}$, as the right-handed neutrino (RHN) channel is sub-dominant. For successful leptogenesis, the inflaton decay to right-handed neutrinos must be kinematically allowed, i.e., the above equation must satisfy the constraint $m_{I} \geq 2 M_{\nu_1}$.
		
		\subsection{Inflationary Observables}
		
		The slow-roll parameters in Eq. \eqref{eq:slow_roll_1} are modified and can now be expressed in terms of $w$ as
		\begin{eqnarray}\label{eq:epsilon}
			\epsilon(w) &=& \frac{1}{2}m_{P}^{2} \left(\frac{V_E^\prime (w)}{J (w) V_E (w)}\right)^2,\\ \label{eq:eta}
			 \eta(w) &=& m_{P}^{2}\left( \frac{V_E^{\prime \prime} (w)}{J^2 (w) V_E (w)}- \frac{J^{\prime} (w) V_E^\prime (w)}{J^3 (w) V_E (w)}\right) ,  \\ \label{eq:xi}
			\xi^2 (w) &=& m_P^4 \left(\frac{V_E^\prime (w)}{J^4 (w) V_E (w)}\right) \left(\frac{V_E^{\prime \prime \prime} (w)}{V_E (w)} - 3 \frac{J^{\prime} (w) V_E^{\prime \prime} (w)}{J(w) V_E (w)} \right. \nonumber \\
			&+& \left.  3 \frac{J^{\prime 2} (w) V_E^{\prime} (w)}{J^2 (w) V_E (w)} -  \frac{J^{\prime \prime} (w) V_E^{\prime} (w)}{J (w) V_E (w)}\right),
		\end{eqnarray}
		where a prime denotes a derivative with respect to $w$. The tensor-to-scalar ratio $r$, the scalar spectral index $n_s$, and the running of the scalar spectral index $dn_s / d \ln k$ to the first order in the slow-roll approximation are expressed as follows:
		\begin{eqnarray}\label{eq:ttr}
			r &\simeq& 16 \epsilon(w_0) , \\  \label{eq:ns2}
			n_s&\simeq& 1-6 \epsilon(w_0) + 2 \eta(w_0), \\ \label{eq:dns}
			\frac{d n_s}{d\ln k} &\simeq& 16\,\epsilon (w_0)\,\eta (w_0) - 24\,\epsilon^2 (w_0) - 2\,\xi^2 (w_0).
		\end{eqnarray}
		The amplitude of the scalar power spectrum is given by 
		\begin{equation}\label{As}
			A_s(k_0)=\left. \frac{1}{{24} \pi^2 m_P^4}\frac{V_E(w)}{\epsilon(w)}\right|_{w(k_0) = w_0},
		\end{equation}
		where $w_0$ is the field value at the pivot scale. The $N_0$ last number of $e$-folds from $w=w_0$ to the end of inflation at $w=w_e$ are expressed as: 
		\begin{equation} \label{N0}
			N_0 = \frac{1}{m_P^2} \int_{w_e}^{w_0}  \dfrac{J^2 (w) V_E (w)}{V_E^\prime (w)} dw,
		\end{equation}
		where the number of $e$-folds $N_0$ are related to the reheat temperature $T_{\text{RH}}$ as given in Eq. \eqref{efolds_tr}.
		
		\subsection{Inflationary Predictions and Comparison with Observations}
		
\begin{figure}[t]
			\centering \includegraphics[width=\textwidth]{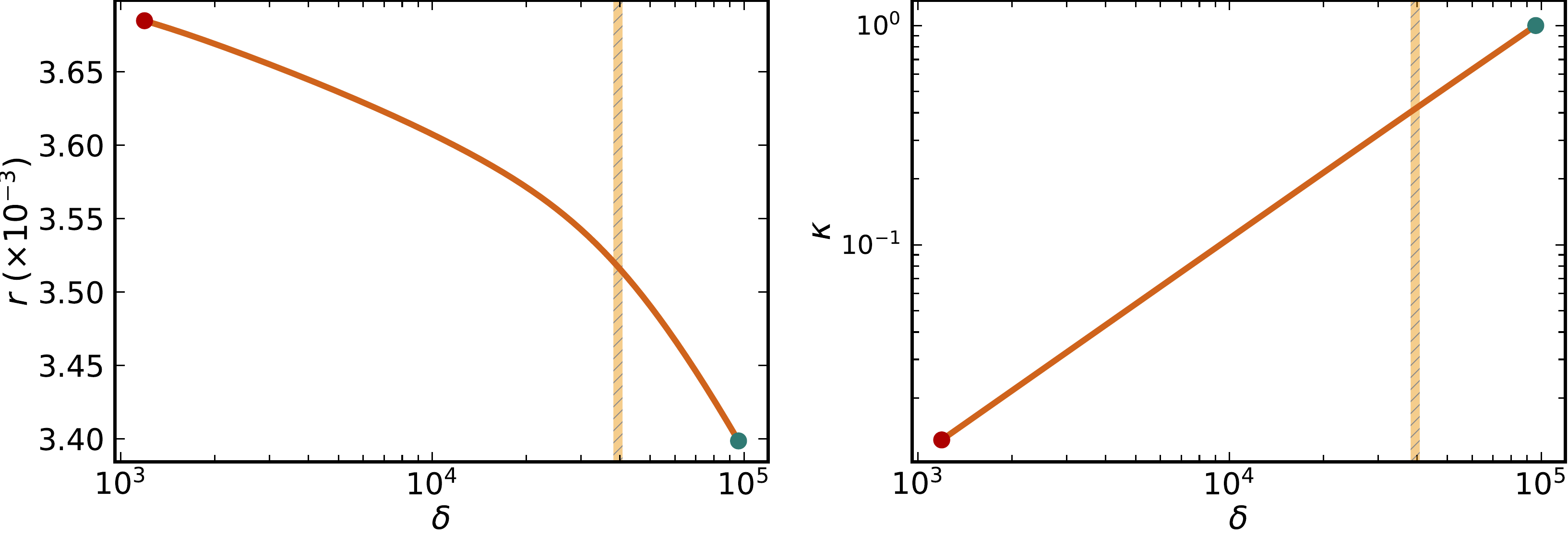}
			\caption{Variation of tensor-to-scalar ratio $r$ (left) and dimensionless coupling $\kappa$ (right) with the non-minimal coupling $\delta$. The red and green circles correspond to $\lvert w_0 \rvert  = m_P$ and $\kappa = 1$ constraints, respectively. The small hatched region represents the parametric space where the kinematic condition ($m_I \geq 2 M_{\nu_1}$) breaks down.}
			\label{fig:nonminimal_higgs_1}
		\end{figure}

\begin{figure}[t]
			\centering \includegraphics[width=\textwidth]{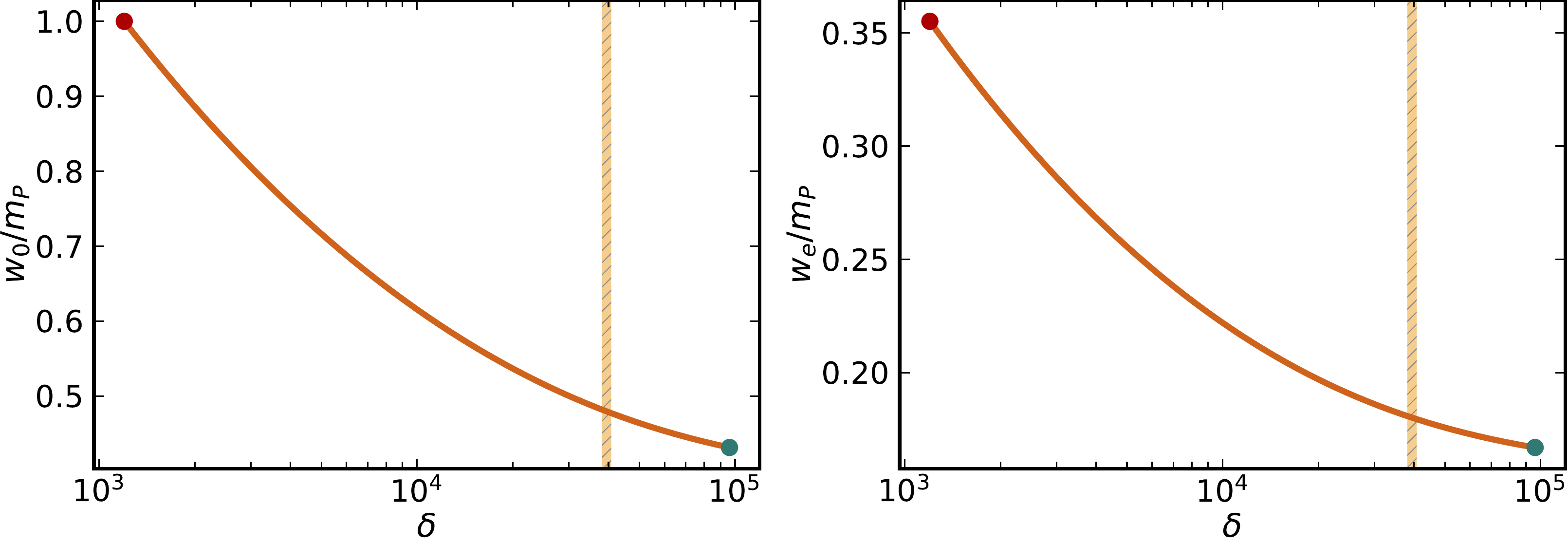}
			\caption{Variation of $w_0/m_P$ (left) and $w_e/m_P$ (right) with the non-minimal coupling $\delta$. The red and green circles correspond to $\lvert w_0 \rvert  = m_P$ and $\kappa = 1$ constraints, respectively. The small hatched region represents the parametric space where the kinematic condition ($m_I \geq 2 M_{\nu_1}$) breaks down.}
			\label{fig:nonminimal_higgs_2}
		\end{figure}

\begin{figure}[t]
			\centering \includegraphics[width=\textwidth]{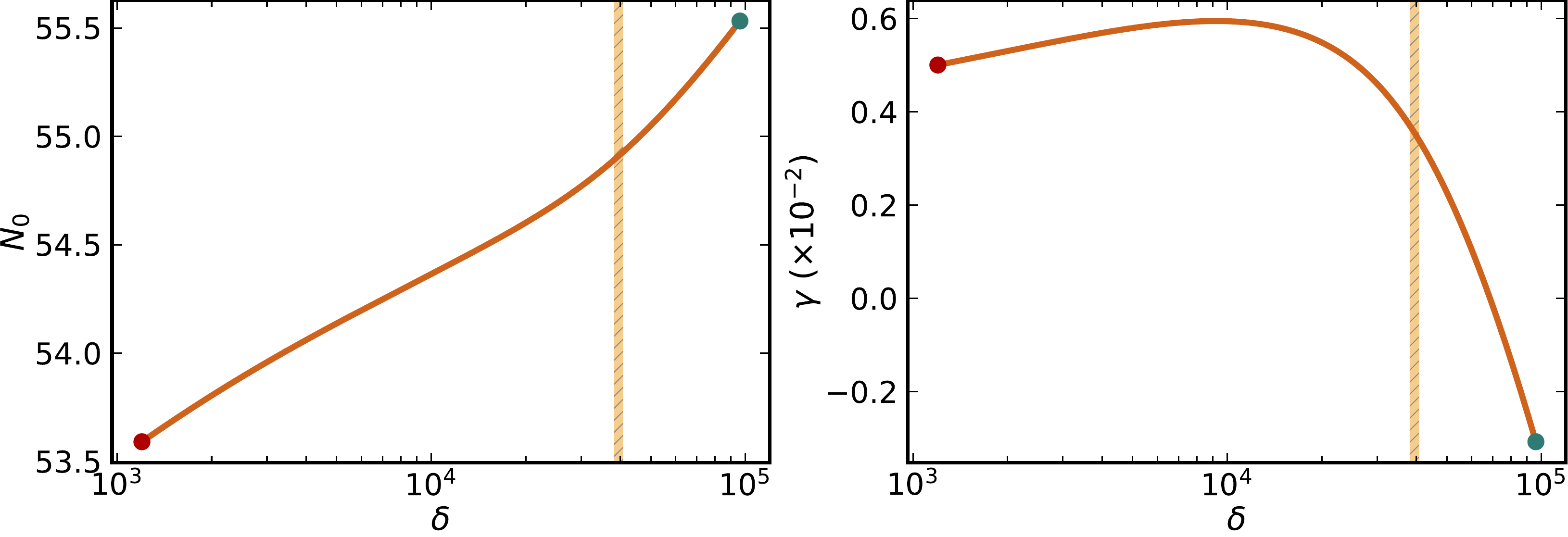}
			\caption{Variation of the last number of $e$-folds $N_0$ (left) and  the non-minimal coupling $\gamma$ (right) with the non-minimal coupling $\delta$. The red and green circles correspond to $\lvert w_0 \rvert  = m_P$ and $\kappa = 1$ constraints, respectively. The small hatched region represents the parametric space where the kinematic condition ($m_I \geq 2 M_{\nu_1}$) breaks down.}
			\label{fig:nonminimal_higgs_3}
		\end{figure}

\begin{figure}[t]
			\centering \includegraphics[width=\textwidth]{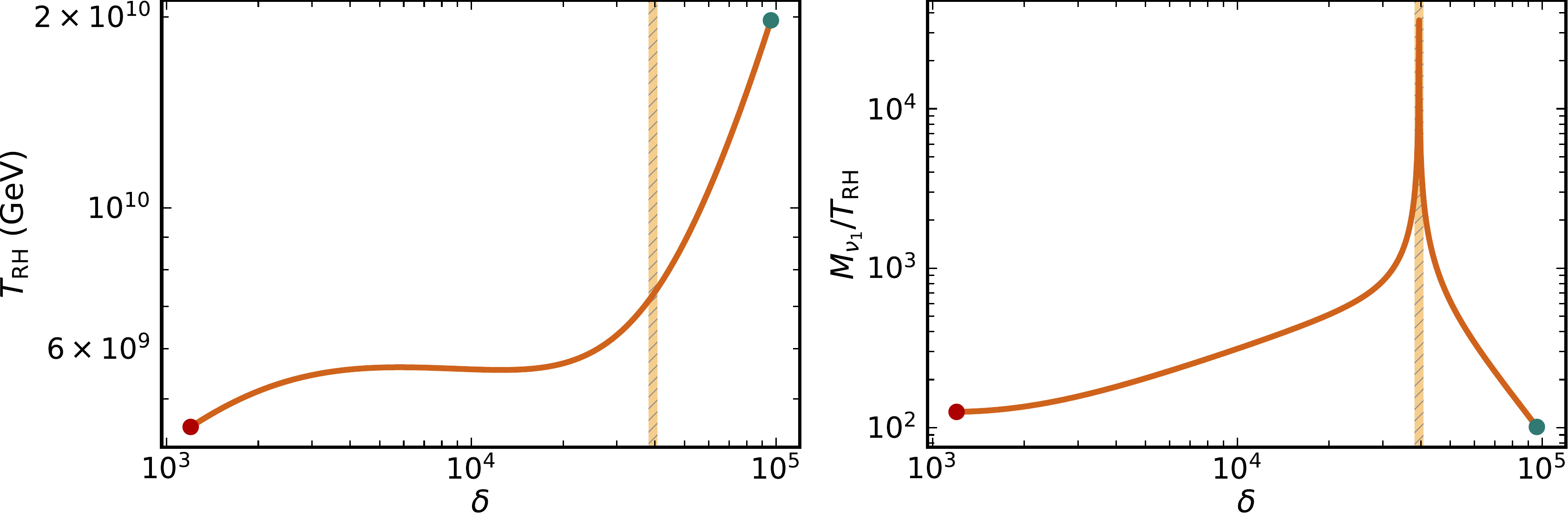}
			\caption{Variation of the reheat temperature $T_{\text{RH}}$ (left) and the ratio $M_{\nu_1}/T_{\text{RH}}$ (right) with the non-minimal coupling $\delta$. The red and green circles correspond to $\lvert w_0 \rvert  = m_P$ and $\kappa = 1$ constraints, respectively. The small hatched region represents the parametric space where the kinematic condition ($m_I \geq 2 M_{\nu_1}$) breaks down.}
			\label{fig:nonminimal_higgs_4}
		\end{figure}

\begin{figure}[t]
			\centering \includegraphics[width=\textwidth]{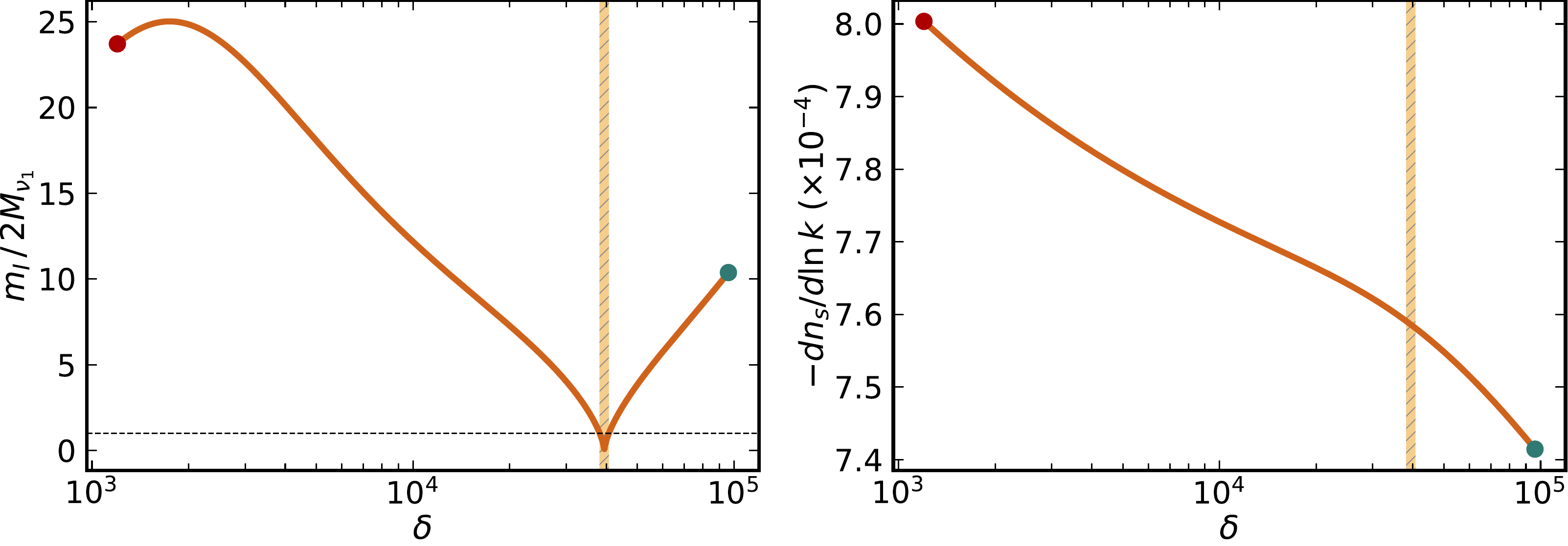}
			\caption{Variation of the ratio $m_I/2 M_{\nu_1}$ (left) and the running of the scalar spectral index $dn_s/d\ln k$ (right) with the non-minimal coupling $\delta$. The red and green circles correspond to $\lvert w_0 \rvert  = m_P$ and $\kappa = 1$ constraints, respectively. The small hatched region represents the parametric space where the kinematic condition ($m_I \geq 2 M_{\nu_1}$) breaks down.}
			\label{fig:nonminimal_higgs_5}
		\end{figure}

		The results of our analysis are depicted in Figs. \ref{fig:nonminimal_higgs_1} - \ref{fig:nonminimal_higgs_5}, where various model parameters and inflationary observables are plotted against the non-minimal coupling $\delta$. Throughout our analysis, we have set the scalar spectral index at the central observationally favored value of the Planck2018 + BICEP2/Keck Array (BK15) bounds ($n_s = 0.9651$) and fixed the mass parameter $M_{*} \simeq 10^{16}$ GeV, which determines $M = \sqrt{M_{*} m_P} \simeq 1.56 \times 10^{17}$ GeV. Additionally, we have ensured that the dimensionless parameter $\kappa \leq 1$ and that the field values remain sub-Planckian ($w_0 \leq m_P$). These constraints are indicated by the green and red circles, respectively. The right hand side of the inequality \eqref{stability_con_NMHI} yields $\simeq 1.4 \times 10^{-5}$. Therefore, we set $\lambda \simeq 1 \times 10^{-5}$ in our numerical calculations.
		
		The hatched region in Figures \ref{fig:nonminimal_higgs_1} - \ref{fig:nonminimal_higgs_5} represents a small region of parametric space where the kinematic condition ($m_I \geq 2 M_{\nu_1}$) breaks down, and as a result, it is excluded from our results. This can be understood from Eq. \eqref{RHN_mass_leptogen}, where it can be seen that the RHN mass $M_{\nu_1}$ depends on the following factor:
		\begin{equation}
			M_{\nu_1} \propto \left(1 + \left( \frac{M}{m_P} \right)^2 - \frac{1}{2} \left(3\delta+2\gamma\right) \left( \frac{M}{m_P} \right)^4\right)^{-2/3},
		\end{equation}
		which, for the coupling values $38111 \lesssim \delta \lesssim 40782$, approaches $0$, leading to large $M_{\nu_1}$. Consequently, $M_{\nu_1}$ becomes greater than the inflaton mass $m_I$, breaking the kinematic condition, as depicted in the left panel of Fig. \ref{fig:nonminimal_higgs_5}. Outside this range of $\delta$, the kinematic bound is easily satisfied. Note that the inflaton mass remains almost constant $m_I \simeq (2.7 - 4.2) \times 10^{13} \text{ GeV} $ for the entire range of $\delta$.
		
		The variation of the tensor-to-scalar ratio $r$ and the coupling $\kappa$ is illustrated in Fig. \ref{fig:nonminimal_higgs_1}, while Fig. \ref{fig:nonminimal_higgs_2} depicts the behavior of $w_0$ and $w_e$ with respect to the non-minimal coupling $\delta$. It can be seen that observable values of the tensor-to-scalar ratio are achieved in the range $3.37 \times 10^{-3} \lesssim r \lesssim 3.86 \times 10^{-3}$ across the entire range of $\delta$. It is noteworthy that the field $w_0$ varies inversely with $\kappa$ due to their inverse relationship: $w_0 \propto 1/\kappa^{1/4}$, as shown below. Fig. \ref{fig:nonminimal_higgs_3} shows the variation of the number of $e$-folds and the non-minimal coupling $\gamma$ with $\delta$. The number of $e$-folds varies slightly, ranging from $53.5 \lesssim N_0 \lesssim 55.5$, while the coupling $\gamma$ assumes very small values and does not affect the inflationary predictions. The variation of the reheat temperature is presented in Fig. \ref{fig:nonminimal_higgs_4}, along with the ratio $M_{\nu_1}/T_{\text{RH}}$, indicating that the out-of-equilibrium condition for leptogenesis is easily satisfied. Finally, Fig. \ref{fig:nonminimal_higgs_4} displays the running of the scalar spectral index, which ranges from $-0.0008 \lesssim dn_s / d\ln k \lesssim 0.00074$, consistent with the assumptions of the $\Lambda$CDM$+r$ model.
		
		It is useful to provide analytical estimate of the values obtained in our numerical results. Using Eq. \eqref{eq:eta} with $\epsilon(w_e) = 1$ and Eq. \eqref{As} with Planck normalization constraint $A_{s} = 2.1 \times 10^{-9}$, we can express $w_0$ and $w_e$ in terms of the non-minimal coupling $\delta$ and $M$ as
		\begin{eqnarray}\label{w0_analytic}
			\frac{w_0}{2 m_P} &\simeq& \left( \frac{2^{5/2} \pi \sqrt{A_s}}{\kappa} \left( 1 + \delta \left(\frac{M}{m_P}\right)^4\right) \right)^{1/4}, \\\label{we_analytic}
			\frac{w_e}{2 m_P} &\simeq& \left( \frac{2}{\sqrt{3} \delta} + \frac{13}{6}  \left(\frac{M}{m_P}\right)^4 \right)^{1/4}.
		\end{eqnarray}  
		Moreover, using Eqs. \eqref{N0}, \eqref{eq:ttr}, \eqref{eq:ns2} and \eqref{eq:dns}, we obtain the following approximate expressions for the number of $e$-folds $N_0$, tensor-to-scalar ratio $r$, scalar spectral index $n_s$ and the running of scalar spectral index $d n_s/d \ln k$:
		\begin{eqnarray}\label{N0_analytic}
			N_0 &\simeq& \frac{3}{4} \left( \frac{\left(\frac{w_0}{2 m_P}\right)^4 - \left(\frac{w_e}{2 m_P}\right)^4}{\frac{1}{\delta} + \left(\frac{M}{m_P}\right)^4}  \right), \\ \label{tts_analytic}
			r &\simeq& \frac{64}{3} \left(\frac{1}{\delta} + \left(\frac{M}{m_P}\right)^4\right)^2 \left( \left(\frac{w_0}{2 m_P}\right)^4 - \left(\frac{M}{m_P}\right)^4\right)^{-2}, \\ \label{ns_analytic}
			n_s &\simeq& 1 - \frac{8}{3}  \left(\frac{1}{\delta} + \left(\frac{M}{m_P}\right)^4\right) \left( \left(\frac{w_0}{2 m_P}\right)^4 + \left(\frac{M}{m_P}\right)^4\right) \left( \left(\frac{w_0}{2 m_P}\right)^4 - \left(\frac{M}{m_P}\right)^4\right)^{-2}, \\ \label{dns_analytic}
			\frac{d n_s}{d\ln k} &\simeq& - \frac{14}{3} \left(\frac{2 m_P}{w_0}\right)^8 \left(\frac{1}{\delta} + \left(\frac{M}{m_P}\right)^4\right)^2. 
		\end{eqnarray}
		Eliminating $w_0$ from Eqs. \eqref{w0_analytic} and \eqref{N0_analytic}, we obtain an explicit relationship between $\kappa$ and $\delta$
		\begin{equation}
			\kappa \simeq \frac{3 \sqrt{2} \pi A_s \delta}{N_0},
		\end{equation}
		 which explains their behavior depicted in the right panel of Fig. \ref{fig:nonminimal_higgs_1}.
		Next, we derive the expressions for $\phi_0$, $n_s$, $r$, and $d n_s/d \ln k$ as functions of the number of $e$-folds $N_0$. Neglecting $w_e$ in Eq. \eqref{N0_analytic}, we express $w_0$ in terms of $N_0$:
		\begin{equation}
			\frac{w_0}{2 m_P} \simeq \left( \frac{4}{3} N_0 \left(\frac{1}{\delta} + \left(\frac{M}{m_P}\right)^4\right) \right)^{1/4}.
		\end{equation}
		Using this expression for $w_0$ in Eqs. \eqref{tts_analytic} - \eqref{dns_analytic} yields:
		\begin{equation}\label{ttr_ns_dns_N0}
			r \simeq \frac{12}{N_0^2}, \quad 
			n_s \simeq 1 - \frac{2}{N_0}, \quad
			\frac{d n_s}{d\ln k} \simeq - \frac{21}{8 N_0^2}.
		\end{equation}
		 It can be checked that for $N_0 \simeq 55$, these expressions yield $r \sim 0.004$, $n_s \sim 0.964$ and $d n_s /d \ln k \sim -0.0008$. These values closely align with the more precise results displayed in Figs. \ref{fig:nonminimal_higgs_1} - \ref{fig:nonminimal_higgs_5}. The MSSM $\mu$ term of $\mathcal{O} (1)$ TeV, can be obtained from Eq. \eqref{mu_term_NMHI} with $m_{3/2} \simeq 100$ TeV, $\kappa \simeq 0.013$, $\delta \simeq 25000$ and $\lambda \simeq 10^{-5}$.

		In summary, for non-minimal coupling $1200 \lesssim \delta \lesssim 96000$, we obtain $0.013 \lesssim \kappa \lesssim 1$, number of $e$-folds $53.5 \lesssim N_0 \lesssim 55.5$, and the RHN mass in the range $5.6 \times 10^{11} \text{ GeV} \lesssim M_{\nu_1} \lesssim 1.6 \times 10^{14} \text{ GeV}$ with the coupling $5.7 \times 10^{-5} \lesssim \alpha_1 \lesssim  2 \times 10^{-2}$. The scalar spectral index closely aligns with the central value of Planck's bounds along with observable values of the tensor-to-scalar ratio $r \sim 3 \times 10^{-3}$. The non-minimal coupling $\delta$ takes on large values\footnote{The standard non-minimal Higgs inflation model encounters an inconsistency \cite{alex_azadeh:2014, barbon_prd79:2009, Burgess_jhep07:2010, Bezrukov_jhep01:2011} related to the validity of the effective field theory. This issue arises because achieving inflation with subplanckian field values $\phi \leq m_P$ requires large values of the coupling $c_{\mathcal{R}}$ ($\delta$ here). Consequently, the inflationary scale becomes larger than the Ultraviolet (UV) cut-off of the effective theory, causing the theory to break down above it. In the model studied here, however, the coupling function has the form $\simeq 1 + c_{\mathcal{R}} \phi^4$, and as shown in \cite{kpallis_plb789:2019, kpallis_jcap03:2015, kpallis_epjc78:2018}, there is no problem with unitarity.} ($\sim 10^5$), a characteristic feature of non-minimal Higgs inflation models.

		\section{Observable Primordial Gravitational Waves} \label{sec5}
		
		The distinctive signature of the tensor signal primarily manifests in the degree scales of the CMB B-mode polarization. Numerous experiments have already been conducted, and others are currently underway, aimed at measuring the B-mode power. The tensor-to-scalar ratio "$r$," which parameterizes the amplitude of primordial gravitational waves, is poised to be measured with significant precision by numerous forthcoming experiments. These experiments encompass a range of missions, including the Simons Observatory \cite{SimonsObservatory:2018koc}, which aims to achieve a measurement of $r$ with $\delta r = 0.003$. CMB-S4 (Cosmic Microwave Background Stage 4) \cite{Abazajian:2019eic, Belkner:2023cmbs4}, on the other hand, seeks to detect $r \gtrsim 0.003$ at a significance level exceeding 5$\sigma$, or establish an upper limit of $r < 0.001$ with 95$\%$ confidence level in the absence of a detection. LiteBIRD (Light satellite for the studies of B-mode polarization and Inflation from cosmic background Radiation Detection) \cite{Allys:2023aja} is dedicated to achieving a measurement of $r$ within an uncertainty of $\delta r = 10^{-3}$. Similarly, PICO (Probe of Inflation and Cosmic Origins) \cite{Aurlien:2023rhd, Hanany:2019} aims to detect $r = 5 \times 10^{-4}$ at a significance level of 5$\sigma$. Additional missions include CORE (Cosmic Origins Explorer) \cite{Finelli:2016cyd}, which is projected to exhibit sensitivity to $r$ as low as $10^{-3}$; PIXIE (Primordial Inflation Explorer) \cite{Kogut:2011xw}, anticipates to measure $r < 10^{-3}$ at the 5$\sigma$ level; and PRISM (Polarized Radiation Imaging and Spectroscopy Mission) \cite{Andre:2013afa}, which aspires to achieve detection of $r$ as low as $5 \times 10^{-4}$ at 5$\sigma$. The large tensor modes ($r \sim 10^{-3}$) obtained in both models of smooth $\mu$-hybrid inflation and non-minimal Higgs inflation are potentially measurable by these upcoming CMB experiments.
		
		\section{Proton Decay} \label{sec6}
		In the $G_{422}$ model, proton decay is mediated by color triplets in $F, F^c \supset d, d^c$, $H^c, \bar{H}^c \supset d_H^c, \bar{d}^c_H$, and $G \supset g, g^c$, rather than via gauge bosons \cite{mansoor:2020pd422}. These color triplets acquire superheavy masses from the following superpotential terms:
		\begin{equation}
			W \supset a\,G H^c H^c + b\,G \bar{H}^c \bar{H}^c \supset m_{d_H^c} g d_H^c +  m_{\bar{d}_H^c} g^c \bar{d}_H^c,
		\end{equation}
		where $m_{d_H^c} = a M$ and $m_{\bar{d}_H^c} = b M$. The $U(1)_R$ symmetry plays a vital role in forbidding undesirable baryon number-violating operators that can lead to fast proton decay. These include $d=4$, $B$ and $L$ violating operators such as:
		\begin{equation}
			\frac{F_i F_j F_k^c H^c}{m_P} \supset \frac{M}{m_P} \left(L_i L_j e^c_k + Q_i L_j d_k^c\right) , \qquad  \frac{F_i^c F_j^c F_k^c H^c}{m_P} \supset \frac{M}{m_P} u_i^c d_j^c d^c_k ,
		\end{equation}
		which are invariant under the gauge symmetry $G_{422}$ but not under the $U(1)_R$ symmetry. Similarly, some gauge invariant $d=5$ proton decay operators can appear at the non-renormalizable level, such as:
		\begin{equation}
			\frac{F_i^c F_j^c F_k^c F_l^c}{m_P} \supset \frac{\left(u_i^c u_j^c d_k^c e_k^c +u_i^c d_j^c d_k^c \nu_k^c\right)}{m_P}  , \qquad  \frac{F_i F_j F_k F_l}{m_P} \supset \frac{Q_i Q_j Q_k L_l}{m_P}  ,
		\end{equation}
		but are also not allowed by the $U(1)_R$ symmetry. The proton decay rate via dimension-6 operators \cite{mansoor:2020pd422} with $M \simeq 2 \times 10^{17}$ GeV, $M_{\nu_i} \simeq 10^{14}$ GeV and natural values of the couplings $a, b \sim 1$, is heavily suppressed to have any observable signature in the next generation of experiments such as Hyper-Kamiokande \cite{Hyper-Kamiokande:2018ofw}, JUNO \cite{Fengpeng:2015juno} and DUNE \cite{DUNE:2015lol, DUNE:2020ypp}. 
		
		\section{Summary} \label{sec7}
		
		We have implemented two inflationary models within the framework of $G_{422} \equiv SU(4)_{C}\times SU(2)_{L}\times SU(2)_{R}$ GUT gauge symmetry, including a smooth variant of $\mu$-hybrid inflation and a non-minimal Higgs inflation model. Both effectively generate the MSSM $\mu$-term, and since the $G_{422}$ GUT gauge symmetry is spontaneously broken to SM gauge group during inflation, the primordial magnetic monopoles are inflated away. Notably, both models exhibit a tree-level inclination, obviating the need for radiative corrections.
		
		The smooth $\mu$-hybrid model, with a minimal canonical K\"ahler potential and soft SUSY masses $m_{3/2} \simeq 100$ TeV, proves incompatible with experimental observations. However, a non-minimal K\"ahler potential enhances the parametric space, resulting in a scalar spectral index $n_s$ that aligns perfectly with the central value of Planck2018 + BK15 data bounds.
		
		The non-minimal Higgs model also predicts a scalar spectral index $n_s$ that aligns very closely to the central observationally favored value of the Planck2018 + BK15 data bounds. In both models, the inflationary phase is followed by reheating, during which the inflaton decays into the lightest right-handed neutrino, leading to non-thermal leptogenesis through the subsequent decay of the right-handed neutrinos. Furthermore, both models yield large tensor modes $r \, (\sim10^{-3})$, potentially measurable in the upcoming generation of CMB experiments.
				
		\acknowledgments{The author extends gratitude to Mansoor Ur Rehman and Constantinos Pallis for their insightful discussions and careful review of the manuscript. Additionally, the author would like to thank the editor for their invaluable suggestions, which greatly enhanced the quality of the manuscript.}

	\end{document}